\documentclass{IEEEoj}
\usepackage{cite}
\usepackage{amsmath,amssymb,amsfonts}
\usepackage{amsthm}
\usepackage{mathrsfs}
\usepackage{algpseudocode}
\usepackage{algorithm}
\usepackage{graphicx,color}
\usepackage{textcomp}
\usepackage{multirow}
\usepackage{booktabs}
\usepackage{xcolor}
\usepackage{manyfoot}
\usepackage{listings}
\usepackage{soul}
\usepackage{color}
\usepackage{float}
\theoremstyle{plain}

\def\BibTeX{{\rm B\kern-.05em{\sc i\kern-.025em b}\kern-.08em
    T\kern-.1667em\lower.7ex\hbox{E}\kern-.125emX}}
\AtBeginDocument{\definecolor{ojcolor}{cmyk}{0.93,0.59,0.15,0.02}}
\def\OJlogo{\vspace{-4pt}\hskip-4pt\includegraphics[height=18pt]{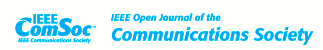}}

\begin{document}
\receiveddate{XX Month, XXXX}
\reviseddate{XX Month, XXXX}
\accepteddate{XX Month, XXXX}
\publisheddate{XX Month, XXXX}
\currentdate{XX Month, XXXX}
\doiinfo{OJCOMS.XXXX.XXXXXX}

\title{A Mobile Agent-Based Hierarchical Reinforcement Learning Framework for Energy-Balanced Data Collection and Wireless Charging in WSNs}

\author{ALI HEIDARIPOUR\IEEEauthorrefmark{1}
AND NASTOOH TAHERI JAVAN\IEEEauthorrefmark{1,2}, SENIOR MEMBER, IEEE}

\affil{Computer Engineering Department, Imam Khomeini International University, Qazvin, Iran}
\affil{Computer Engineering Department, Amirkabir University of Technology, Tehran, Iran}

\corresp{CORRESPONDING AUTHOR: Nastooh Taheri Javan (e-mail: nastooh@eng.ikiu.ac.ir, nastooh@aut.ac.ir).}

\markboth{Mobile Agent-Based HRL for Energy-Balanced WSNs}{Heidaripour \textit{et al.}}

\begin{abstract}
Energy imbalance remains a key challenge in Wireless Sensor Networks (WSNs), as nodes near the base station deplete their energy faster due to heavy forwarding loads. While mobile agents (MAs) have been employed for either data collection or sensor charging, existing approaches lack adaptability and fail to integrate both functions under realistic hardware constraints. This paper introduces a unified mobile agent framework that performs both data collection and wireless charging sequentially under single-antenna limitations. The agent's decision-making is formulated as a two-layer Hierarchical Reinforcement Learning (HRL) problem, where the upper layer optimizes movement planning and the lower layer determines the appropriate service based on real-time network states. This hierarchical structure enables the agent to learn adaptive task scheduling policies without predefined rules. Extensive simulations demonstrate that the proposed method achieves up to 15\% longer network lifetime and more balanced energy distribution compared with state-of-the-art mobile agent and deep RL approaches.
\end{abstract}


\maketitle

\section{Introduction}\label{sec1}

Wireless Sensor Networks (WSNs), composed of small battery-powered nodes that collect and transmit environmental data ~\cite{najjar2025dynamic}, are widely applied in diverse domains such as industry, environment, and healthcare \cite{trigka2025wireless}. They have become fundamental to modern Internet of Things (IoT) applications \cite{11358823}. However, as WSNs expand, new challenges emerge in data transmission, energy efficiency, and long-term network performance \cite{ khan2025integrating}. Due to their limited communication range, sensor nodes typically rely on multi-hop transmission to forward data to the base station \cite{11604084, 11142319}. Consequently, nodes located closer to the base station experience a heavier communication load, as they must not only send their own data but also relay data from other nodes \cite{Salehi2018}. This uneven energy consumption leads to bottlenecks, causing these nodes to deplete their energy rapidly, which in turn shortens the overall network lifetime and may eventually disrupt multi-hop communication links \cite{ZHAO2022102872}. As a result, data transmission is interrupted, and the base station loses visibility into the network’s status.

To balance energy consumption and extend the lifetime of WSNs, researchers have explored various strategies \cite{khan2025integrating, hussain2024comprehensive}. Common approaches include energy harvesting from the environment \cite{11579373}, improving clustering algorithms \cite{11400533}, data aggregation \cite{nandal2025energy}, fixed charging stations \cite{10964260}, and the use of mobile sinks \cite{dholey2023mshrp}. More recently, mobile agents (MAs)—autonomous entities capable of moving within the network—have gained increasing attention due to their flexibility and multifunctional capabilities \cite{temene2022survey, kaur2025mobile}. By traversing the network, mobile agents can collect data directly from sensor nodes and deliver them to the base station, thereby reducing multi-hop communication and alleviating network congestion \cite{S202224}. Furthermore, they can enhance network longevity by performing on-demand tasks or recharging sensor nodes through radio-frequency (RF)-based wireless energy transfer \cite{charger}.

In previous research on mobile agents in WSNs, many studies have focused solely on either data collection \cite{tran2023mobile} or sensor charging \cite{JIANG2024103458}, and only a few have attempted to integrate both tasks into a single mobile agent \cite{8669799}. Even in those few cases, the agent usually follows fixed rules—for example, predefined paths and actions \cite{chen2022mobile}. However, fixed rules make it difficult for mobile agents to adapt their actions—such as choosing when and where to charge or collect data—especially in networks where conditions frequently change. This challenge is further complicated by technical limitations that prevent a mobile agent from performing both data collection and sensor charging simultaneously in practical single-antenna systems \cite{perera2018swipt, ashraf2021swipt}. Attempting simultaneous operations leads to electromagnetic interference and signal separation difficulties \cite{baikova2017interference}. Therefore, practical implementations require the mobile agent to alternate between these tasks sequentially. These limitations highlight the need for an adaptive decision-making framework that intelligently schedules when to perform each task based on network conditions.

To address the limitations of previous rule-based mobile agent strategies, we propose a Hierarchical Reinforcement Learning-based framework, called DCR-HRL (Data Collection, Charging, and Routing with Hierarchical Reinforcement Learning), in which a single mobile agent adaptively alternates between data collection and sensor charging tasks under realistic single-antenna constraints. The proposed model explicitly decomposes the agent’s decision-making into two coordinated layers: (i) a service-selection layer that determines whether to collect data or recharge nodes based on local energy and buffer states, and (ii) a movement-planning layer that selects the next destination considering aggregate energy and buffer conditions of the adjacent subdomains. This hierarchical separation enables the agent to learn optimal sequential policies through experience, without relying on predefined rules or simultaneous transmission assumptions. Unlike existing single-layer or heuristic approaches, DCR-HRL unifies data collection, energy replenishment, and mobility optimization within a single adaptive learning framework. To the best of our knowledge, this is the first work that integrates both data collection and wireless charging in a single mobile agent through hierarchical decision learning, achieving significant improvements in energy balance and network lifetime over existing methods.

The behavior of the mobile agent is formulated as a Markov Decision Process (MDP) \cite{puterman1990markov} divided into two sub-problems and solved using a Hierarchical Q-learning algorithm \cite{chen2011HQL}. At the higher layer, the agent learns to choose the optimal path through the network based on the buffer occupancy and node energy levels of the candidate destination subdomains. At the lower layer, after reaching a chosen location, the agent decides whether to collect data or charge surrounding sensors depending on their real-time needs. This decision-making cycle is repeated continuously.

The overall operation of the proposed hierarchical framework can be intuitively described as follows. As the mobile agent interacts with the network, it gradually learns to make better decisions for both task scheduling and movement patterns. Over time, it becomes more effective in delivering services and managing energy consumption. By intelligently scheduling these two tasks in sequence based on network dynamics, the agent maintains balanced energy distribution across the network while adhering to the constraint that only one task can be performed at a time.

To evaluate the effectiveness of the proposed method, extensive simulations were conducted and compared with several recent mobile agent-based approaches. The results show that the proposed solution increases the network lifetime by an average of 15\% compared to existing methods.

The main contributions of this paper include:
\begin{itemize}
    \item Proposing a unified framework that enables a single mobile agent to adaptively alternate between data collection and sensor charging tasks in wireless sensor networks, explicitly addressing single-antenna constraints that prevent simultaneous operations in networks.
    \item Formulating the mobile agent's decision-making process as two interdependent MDP sub-problems (service selection and movement planning) to enable hierarchical decision coordination in grid-based deployments.    
    \item Developing a model-free hierarchical reinforcement learning solution based on Q-learning that jointly optimizes service scheduling and routing without explicit transition models.    
    \item Demonstrating the robustness of the proposed framework across diverse network scenarios through comprehensive simulations against both learning-based and deterministic baselines.
   
\end{itemize}

The remainder of this paper is organized as follows: Section II outlines the motivation behind the study. Section III reviews related work on the use of mobile agents in WSNs. Section IV describes the system model. Section V details the proposed approach and presents the analytical models. Section VI evaluates the performance of the method through simulations and discusses the results. Finally, Section VII concludes the paper and outlines future research directions.

\section{Preliminary Exploration and Motivation}\label{sec2}

\begin{figure}[h!]
    \centering
    \includegraphics[width=0.50\textwidth]{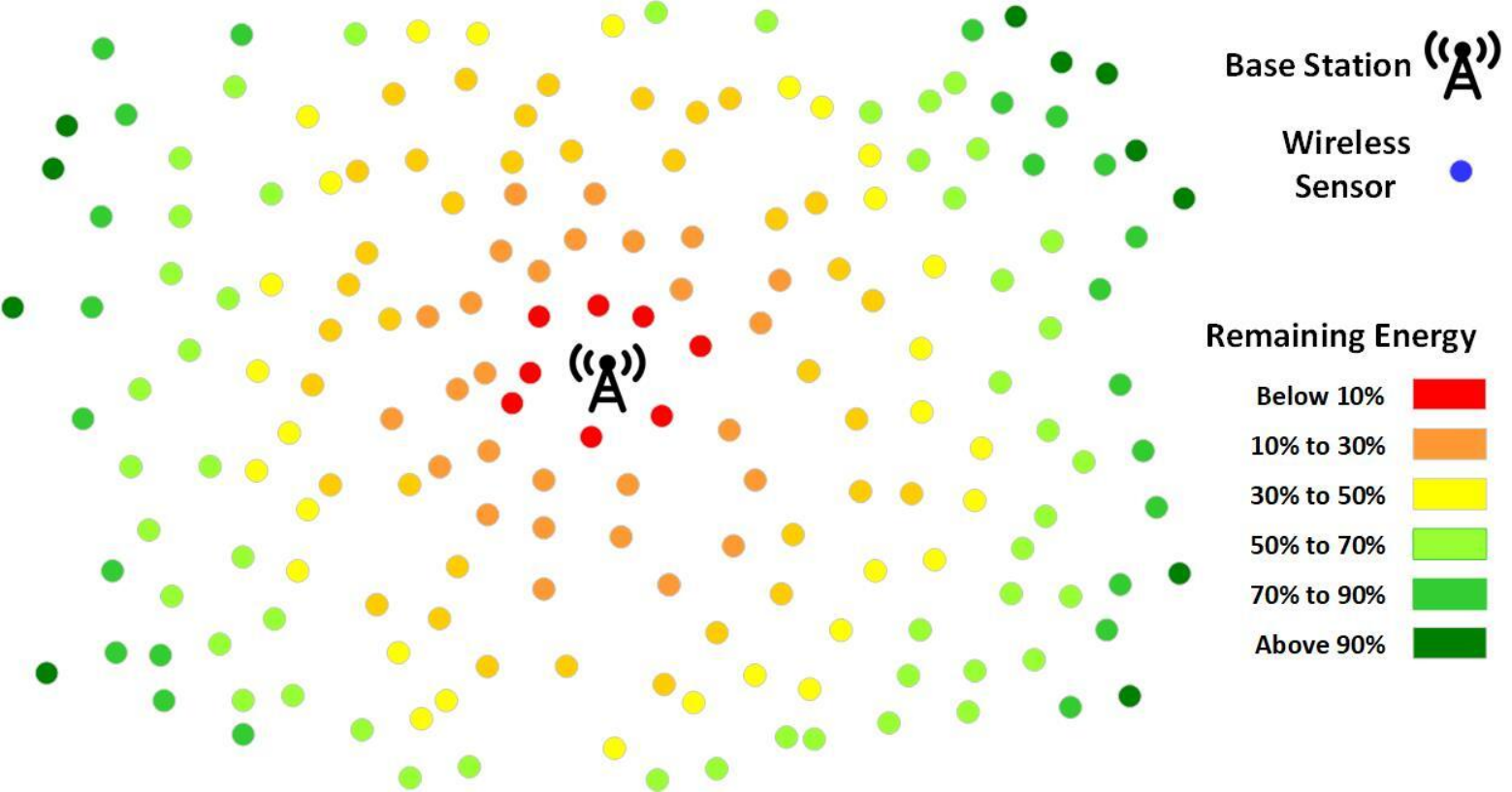} 
    \caption{Remaining energy distribution across sensor nodes after multi-hop data transmission. Nodes near the base station (red/orange) deplete their energy significantly faster than distant nodes (green) due to heavy forwarding loads, illustrating the energy hole problem in WSNs.}
    \label{fig:network_model}
\end{figure}

To highlight the impact of unbalanced energy consumption in Wireless Sensor Networks, we implemented a simple multi-hop data transmission model in a simulation tool. In this simulation, 1,000 fixed nodes were randomly placed in a 300 × 400-meter area. Each node has a 50-meter transmission range and continuously sends data packets to the base station at a fixed rate. Based on the network topology, sensors transmit their data to the base station using multi-hop communication. Figure \ref{fig:network_model} illustrates the average remaining energy of sensors after the simulation. Each data point in the figure represents the average energy of 5–7 nearby nodes, averaged across multiple runs to minimize random bias.

As shown in Figure \ref{fig:network_model}, sensors located closer to the base station deplete their energy considerably faster than distant nodes, because they are responsible for forwarding both their own data and the traffic of others. The difference in average remaining energy between inner and outer regions exceeds 40\%, leading to early node failures and reduced overall network lifetime. In this study, network lifetime is defined as the time elapsed until the first node exhausts its energy. These results confirm the well-known “energy hole problem” in WSNs \cite{dogra2021novel}, where nodes near the sink suffer from disproportionate energy consumption due to multi-hop forwarding.

A potential approach to mitigate this imbalance is the use of a mobile agent (MA) that traverses the network to assist nodes. The simulation results indicate that a mobile agent can help extend network lifetime through two complementary mechanisms. First, it can collect data directly from remote sensors and deliver them to the base station, thereby reducing multi-hop transmissions and relieving heavily loaded nodes near the sink. Second, the agent can recharge nodes with high energy consumption using wireless energy transfer. This combination helps distribute energy usage more evenly and prolong the operational period of the network. These observations motivate the design of an adaptive mobile agent capable of dynamically deciding when to collect data and when to recharge nodes according to the network’s real-time conditions.

However, practical implementations face critical hardware limitations. Current single-antenna architectures do not support Simultaneous Wireless Information and Power Transfer (SWIPT) \cite{perera2018swipt} due to electromagnetic interference between energy and data signals. Studies have shown that signal separation in such configurations is highly inefficient and increases system complexity and cost \cite{ashraf2021swipt, baikova2017interference}. Consequently, a mobile agent cannot perform both data collection and charging concurrently; instead, these operations must be executed sequentially. While using multiple antennas could theoretically enable simultaneous operation, it significantly increases hardware cost and power consumption, making it impractical for most real-world WSN deployments. These constraints make sequential and adaptive task scheduling not only algorithmically relevant but also physically realistic.

Even when a mobile agent uses separate antennas for wireless charging and data collection, the two operations can still interfere with each other. Energy transmission can generate interference to nearby information receivers operating in the same frequency band, motivating a joint design of energy and data transmission; the general asymmetry in receiver sensitivities also means that energy-transfer interference tends to be substantially stronger than the information signal itself, a coexistence challenge noted as an open problem for wireless powered communication systems \cite{7081084}. This effect has been experimentally confirmed in a WSN setting: when a wireless charging source and a data transceiver operate simultaneously within the same ISM band, the charging signal's high transmit power generates harmonics that degrade the signal-to-noise ratio of the much weaker data signal, disrupting communication; alternating rather than simultaneous operation was found to be necessary as a result \cite{s19153330}. The complete isolation of two antennas requires their simultaneous self-interference cancelation on all levels of the chain: antenna, analog and digital, and the total attenuation is required to be up to 110 dB, which is fundamentally impractical for the limited and constrained WSN hardware design \cite{s22051699}. Even the specialized engineering work on achieving the simultaneous data and wireless power transmission in sensor networks requires sophisticated double-LCC circuit design, impedance matching and wave filtering, and multi-objective optimization for reducing mutual interference to the acceptable level \cite{10471294}.

Most existing mobile-agent-based strategies rely on predefined movement paths and fixed service rules\cite{chen2022mobile}. The agent’s trajectory, stopping points, and task types are typically determined in advance, limiting their adaptability to changing network conditions. As a result, their effectiveness in balancing energy consumption and extending network lifetime remains limited \cite{Xie}. These limitations underscore the need for a learning-based and context-aware framework that enables the mobile agent to make intelligent, real-time decisions about service scheduling and movement planning based on the actual state of the network.

At the core of this study lies an HRL strategy designed to address these challenges. In the proposed approach, the mobile agent learns to make adaptive and efficient decisions for both task scheduling and movement. As it navigates through the network, the agent selects stopping points and determines, based on the state of nearby sensors, whether to perform data collection or energy recharging—two tasks that are executed sequentially due to single-antenna constraints. By continuously interacting with the environment, the agent gradually refines its decision policy, optimizing both the order of actions and its movement trajectory. This adaptive behavior allows the agent to reduce multi-hop traffic, maintain balanced energy consumption, and extend the overall network lifetime. The detailed formulation and learning process are presented in the following sections.

\section{Related Works}\label{sec4}
As mentioned earlier, many studies have focused on distributing energy consumption in sensor networks to improve network lifetime. Some of these approaches include clustering-based methods \cite{Asha} and data-gathering techniques \cite{Xie}. On the other hand, the idea of using a mobile agent has also been explored in previous research to some extent. In some mobile agent-based approaches, the agent's sole task is data collection from across the network. In others, the mobile agent acts as a moving charger, supplying power to sensor nodes. However, only a few studies have considered integrating both tasks into a single mobile agent. In the following, we review the key mobile agent-based approaches proposed in the literature and discuss their focus and limitations.

According to \cite{He2013OndemandCI}, Mobile Charging Technologies (MCTs) are classified into periodic and on-demand charging strategies. These methods are reviewed based on design attributes such as energy efficiency, constraints, and performance metrics. Additionally, the paper outlines recent advancements in wireless power transfer technologies and highlights the challenges associated with extending the lifetime of sensor networks, thereby serving as a reference for developing mobile charging solutions.

Based on \cite{wang2019}, an intelligent data gathering mobile agent-based scheme, IDGS-DF, is proposed for WSNs. The approach utilizes a neural network for data fusion, partitioning the sensor field into grids, with cluster heads selected based on a neural network-derived score. A mobile agent collects data from predefined paths to optimize energy efficiency and extend network lifetime. This method reduces redundant transmissions, lowers energy consumption, and establishes an effective framework for distributed sensor networks.

According to \cite{Abdelrahman}, an adaptive dynamic clustering method, DCMDC, is introduced to optimize Mobile Data Collectors (MDCs) in WSNs. By organizing the network into Service Zones (SZs), the approach reduces signaling overhead, improves routing efficiency, and balances energy consumption. This enhances data delivery, minimizes delays, and extends network lifetime, making it effective for dynamic WSN environments.

According to \cite{Tian}, the authors suggest enhancing the energy efficiency of rechargeable wireless sensor networks by leveraging a mobile wireless charging vehicle (WCV) exclusively for the purpose of charging. Through optimization of the WCV's travel distance to sensor nodes and via a nearest-neighbor algorithm determined by Hamiltonian cycles, the authors offer the minimization of overall charging time. A dynamic node replacement strategy is also provided to prevent coverage holes due to dead nodes. This research is only focused on mobile charging and not on data collection, so it will be an authentic source to compare with systems that perform dual functions.

In this respect, the approach \cite{DONG201458} invokes mobile agents for efficient data collection in IoT-enabled WSNs. While navigating around the network, these mobile agents aggregate information to avoid redundant transmission and waste of energy associated with data aggregation. The authors focus on optimization paths of the agent to minimize overall energy utilization while ensuring timely data aggregation. The proposed approach extends the network's operational lifetime by dynamically adapting to network conditions such as node density and traffic patterns. This highlights the versatility of mobile agents in mitigating scalability and energy constraints in various IoT scenarios.

According to \cite{tran2023mobile}, the use of mobile agents in ZigBee-based WSNs for data gathering is analyzed. The simulation results of various routing algorithms demonstrate the potential for energy savings with mobile agents. Current research extensively focuses on low-energy, long-range communication using ZigBee, presenting it as a highly feasible technology for applications requiring continuous data updates. This architecture enhances scalability by ensuring reliable data aggregation in densely deployed sensor networks.

According to \cite{Xie}, a scheme is proposed in which a single mobile platform is utilized for both Wireless Energy Transfer (WET) and data collection in sensor networks. This work introduces a near-optimal scheme aimed at determining the most efficient scheduling of the mobile platform to address both energy refilling and data collection tasks.

According to \cite{Qadori}, a data collection mechanism for wireless sensor networks is proposed, utilizing single-agent and multi-agent itinerary planning algorithms to enhance energy efficiency. A key contribution of this work is the optimization of mobile agent route planning to minimize energy consumption during data collection.

According to \cite{chen2022mobile}, a novel mobile charging strategy for wireless rechargeable sensor networks (WRSNs) is proposed. The authors introduce a dual-side charging mechanism in which a mobile charging robot (MR) traverses a pre-planned path to charge sensors. The MR’s path is derived from a power diagram that accounts for the remaining energy of sensors and the distances between them, ensuring that neighboring nodes on both sides are charged while their data is gathered. To further enhance performance, a clustering-based approach is integrated into the path planning, reducing the total travel distance, energy consumption, and overall task completion time. Simulation results indicate that this integrated strategy significantly outperforms baseline methods in terms of energy saving, distance reduction, and time efficiency.

More recently, \cite{s25030956} propose a joint energy replenishment and data collection scheme for WRSNs using two types of mobile chargers—one that charges designated data storage nodes while collecting their cached data, and another that charges the remaining nodes—optimized via mean-shift clustering and a hybrid particle swarm algorithm to reduce charging and collection delay. On the data collection side, \cite{https://doi.org/10.1002/dac.70199} propose an energy-aware framework that places anchor points using a weighted set cover approach refined by fuzzy C-means clustering, with routing via minimum spanning tree and Euler graph analysis, reducing the number of required collectors and travel distance. Unlike DCR-HRL, both approaches rely on multiple specialized mobile agents and classical optimization rather than a single adaptive agent, and neither addresses the single-antenna constraint that motivates our sequential task scheduling.

A related direction is the exploration of methods by which edge intelligence systems could be sped up by using Integrated Sensing and Communication (ISAC). As per \cite{10005142}, traditional edge intelligence systems perform their sensing and communication processes serially, which results in an excessive creation and upload of datasets. In order to tackle the issue, it merges both processes together such that the wireless signal is used to not only sense data but also to upload it while simultaneously optimizing resource allocation and beamforming for ISAC gains. Our DCR-HRL approach can be extended or compared to this approach of edge intelligence system design, especially in terms of joint optimization of communication, computation, and mobility. However, while ISAC deals with the acceleration of edge-based computations within cellular networks, our framework focuses on energy balance and sequential task processing in WSNs with the constraint of single antenna usage. The common point in both approaches is that they both consider the issue of the joint optimization of resource allocation within certain limitations (sensing and communication in the case of ISAC, charging and data collection in the case of WSNs).

Apart from the ground-based mobile agents, another relevant category of mobile data gathering systems based on UAVs has been introduced. For instance, \cite{10273835} proposes a UAV-assisted system which optimizes trajectory, localization accuracy, bandwidth, and beamwidth using the Age of Information criteria in order to ensure data freshness, resulting in about 50\% improvement in performance over communication-only designs. Another example is \cite{11177503} which introduces a satellite-assisted UAV control architecture that co-optimizes sensing schedule, power management, and trajectory tracking for improving the energy efficiency of IoT remote data gathering using a deep Q-network based algorithm. While such UAV-assisted methods show the flexibility and energy awareness of mobile data gathering, they rely on aerial UAVs powered by constant power sources and do not consider the co-optimization of wireless charging and data gathering with single-antenna hardware limitations.

Recent work on massive IoT access management also targets similar joint efficiency goals. \cite{11543375} propose a hybrid fuzzy logic-guided random access management scheme that prioritizes IoT device access based on delay, interference, and energy criteria, combining a double-queue preamble/backoff structure with preamble parallelization and access-class-barring optimization to jointly improve throughput and energy efficiency across heterogeneous traffic loads. \cite{10906650} similarly develop a QoS-aware joint access control and resource allocation scheme that derives a closed-form preamble utilization ratio and uses colliding preamble reuse to reduce access delay while improving throughput, at lower computational complexity than existing random access control schemes. Though these works operate at the random access layer rather than on mobile-agent scheduling, they share the same underlying goal pursued in this study: jointly balancing energy, throughput, and delay under resource constraints in IoT networks.

More recently, reinforcement learning has also been applied to other wireless resource-allocation problems. \cite{11319344} proposes a layout-specific channel model for indoor Fluid Antenna Systems (FAS) and employs Group Relative Policy Optimization (GRPO), a policy-optimization-based reinforcement learning algorithm, to jointly optimize antenna positioning, beamforming, and power allocation, outperforming proximal policy optimization (PPO) in sum-rate while achieving substantial computational savings over PPO baselines. This work illustrates the broader applicability of reinforcement learning to wireless system design.

While mobile agents have been used for either data collection or charging in many studies, most existing methods still rely on predefined rules and are unable to adapt to dynamic network conditions. In this paper, we present a different approach that uses reinforcement learning to enable the mobile agent to make intelligent, real-time decisions based on the actual state of the network. This allows the agent to perform data collection and sensor charging sequentially in a more flexible and efficient way. To clarify how our method compares with previous works, Table~\ref{table_comp} provides a summary comparison, highlighting the key features and limitations of existing approaches alongside our proposed solution.

\begin{table*}[h!]
    \centering
    \small
    \caption{Comparison of Mobile Agent-Based Approaches in WSNs}
    \label{table_comp}
    \renewcommand{\arraystretch}{1.1} 
    \setlength{\tabcolsep}{3pt}
    \begin{tabular}{|p{1.4cm}|c|c|p{3cm}|p{3cm}|p{3cm}|}
    \hline
    \textbf{ref.} & \textbf{Charging} & \textbf{Data Collection} & \textbf{Approach} & \textbf{Pros} & \textbf{Cons/Limitations} \\
    \hline
    
    \cite{Xie} & 
    \checkmark & \checkmark & 
    Single platform for WET and data collection & 
    Unified scheduling & 
    Complex scheduling, pre-planned paths \\
    \hline
    
    \cite{wang2019} & 
    & \checkmark & 
    NN-based data fusion with grid partitioning & 
    Reduced transmissions; lower energy & 
    NN training overhead \\
    \hline

    \cite{Abdelrahman} & 
    & \checkmark & 
    Adaptive dynamic clustering (DCMDC) for MDC & 
    Lower overhead; extended lifetime & 
    Complex clustering setup \\
    \hline

    \cite{Tian} &
    \checkmark & & 
    WCV path planning via NN + Hamiltonian &
    Reduced charging time; &
    Limited channel model does not consider real-time dynamics \\

    \hline
    \cite{DONG201458} & 
    & \checkmark & 
    MAs for IoT data aggregation & 
    Reduces redundancy; adaptive & 
    Increased algorithmic complexity \\

    \hline
    \cite{tran2023mobile} & 
    & \checkmark & 
    MAs in ZigBee WSN; low-energy, long-range comm. & 
    Energy saving; scalable & 
    Limited to ZigBee \\

    \hline
    \cite{Qadori} & 
    & \checkmark & 
    Single-/multi-agent itinerary planning & 
    Enhanced energy efficiency & 
    Complex itinerary design \\

    \hline
    \cite{chen2022mobile} & 
    \checkmark & \checkmark & 
    Dual-side charging + clustering + TSP routes & 
    Minimizes travel distance & 
    Implementation complexity \\

    \hline
    \cite{s25030956} & 
    \checkmark & \checkmark & 
    Two-type mobile chargers + mean-shift clustering + PSO & 
    Reduced charging/collection delay & 
    Requires multiple chargers; non-adaptive scheduling \\
    
    \hline
    \cite{https://doi.org/10.1002/dac.70199} & 
    & \checkmark & 
    Weighted set cover + fuzzy C-means + MST/Euler routing & 
    Fewer collectors; shorter travel distance & 
    No charging; static optimization, not adaptive \\

    \hline
    \cite{10005142} & 
    & \checkmark & 
    ISAC-based joint sensing and communication for edge intelligence & 
    Reduced dataset generation and uploading time & 
    Targets edge servers, not WSN energy balancing \\
    \hline
    
    \hline
    \cite{10273835} & 
    & \checkmark & 
    UAV-aided localization and communication with AoI-based scheduling & 
    Joint trajectory and resource optimization; 
    50\% performance gain & 
    No charging; requires stable UAV power; aerial platform only \\

    \hline
    \cite{11177503} & 
    & \checkmark & 
    Satellite-assisted UAV control with DQN-based sensing and communication scheduling & 
    Energy-efficient; handles remote IoT areas & 
    No wireless charging; relies on satellite infrastructure \\

    \hline
    DCR-HRL 
    & \checkmark 
    & \checkmark 
    & RL-based sequential approach for charging, data collection, and routing 
    & Adaptive, dynamic for all tasks 
    &  Increases delay \\
    \hline

    \end{tabular}
\end{table*}

\section{System Model and Assumptions}\label{sec5}

This section presents the system model along with the key assumptions underlying the proposed approach.

\subsection{Network Model}

In this study, the network is divided into several square-shaped subdomains, where each subdomain represents a possible stationing location for the mobile agent. Each subdomain contains \( N \) nodes using uniform distribution to make sure each grid has almost the same number of sensors. The RF capability of the mobile agent operates within a well-defined service radius of $R_{\text{service}} = 10$ meters, allowing it to serve up to $M_{\text{max}} = 5$ nodes simultaneously within each subdomain through broadcast-based energy transfer and data collection protocols. By using a uniform distribution  \cite{jain2023energy}, we ensure that each subdomain contains approximately the same number of sensors. The placement for each sensor is determined in Equation~\eqref{eq:sensor_placement}.
\begin{equation}
y \sim \text{Uniform}(y_{\text{min}},\ y_{\text{max}}),\quad
x \sim \text{Uniform}(x_{\text{min}},\ x_{\text{max}})
\label{eq:sensor_placement}
\end{equation}

Each sensor discovers its neighbors using a communication range of \( R_{\text{comm}} \). At the discovery stage, multi-hop communication is established with the base station so that each sensor can find its neighbors and determine the corresponding distance to the next node. In order to minimize energy consumption, sensors send their data packets to the neighbor within their RF communication range that is closest to the base station, following a greedy approach. Once deployed in the network, the sensors remain static.

Each sensor saves the list of neighbors in its memory in a sorted manner so that if the next hop dies and the connection with the closest node is terminated, a new connection will be established with the second closest node to the base station within the neighbors. In every timestep, each sensor generates between 1 and 3 data packets, depending on the available buffer space. In this case, energy depletion increases as packet generation grows. Each sensor has the capacity to store up to 100 data packets in its buffer.

\subsection{Communication Model}

In this research, the success rate of data transmission depends on several factors. A node can only transmit data if the receiving node (next hop) has enough energy to accept it; otherwise, the packet is dropped. A sender continues to transmit data packets until its buffer becomes empty. If it lacks sufficient energy, the sensor sends packets one by one until it runs out of energy, aiming to make the most of what remains.

Only sensors within the sender’s communication range can receive the packets. The transmission process also stops if the receiving node does not have enough buffer space. Since multi-hop communication consumes a significant amount of energy, direct transmission to the MA—when it is within range—is usually much more efficient.

\subsection{Energy Model}
Each sensor is initialized with an energy \( E_{\text{init}} \). The energy consumed for transmitting a packet, as given in \cite{heinzelman2000}, is calculated in Equation~\eqref{eq:energy_tx}.
\begin{equation}
\label{eq:energy_tx}
E_{\text{tx}}(k, d) = E_{\text{elec}} \times k + E_{\text{amp}} \times k \times d^n
\end{equation}
where \( E_{\text{tx}}(k, d) \) is the energy to transmit a \( k \)-bit packet over a distance \( d \). \( E_{\text{elec}} \) \( \text{nJ/bit} \) is the energy used in the transmitter and receiver circuitry, and \( E_{\text{amp}} \) \( \text{pJ/bit/m}^2 \) is the energy needed for signal amplification. The parameter \( n \) is the path loss exponent, which depends on the propagation environment. In free-space conditions, with no obstructions, the energy loss is a quadratic function (\( n = 2 \)). In obstructed environments, like urban or indoor environments, the signal decays more quickly, and the path loss exponent \( n > 2 \).

Similarly, the energy consumed for receiving a packet, as given in \cite{heinzelman2000}, is calculated in Equation~\eqref{eq:energy_rx}.
\begin{equation}
\label{eq:energy_rx}
E_{\text{rx}}(k) = E_{\text{elec}} \times k
\end{equation}

The energy required to charge a sensor at a distance \( d \) by the mobile agent, as given in \cite{naderi2015}, is calculated in Equation~\eqref{eq:energy_charge}.
\begin{equation}
\label{eq:energy_charge}
E_{\text{charge}} = \eta \times P_{\text{MA}} 
\times \left( \frac{d_0}{d} \right)^n \times \Delta t_{\text{charge}}
\end{equation}
where $\eta$ represents the efficiency factor of energy transfer, indicating that half of the energy effectively reaches the sensors. The term $P_{\text{MA}}$ denotes the rate of energy consumption by the mobile agent during the charging process. The parameter \( d_0 \) is a reference distance, typically assumed to be 1 meter, used to model the path loss in wireless sensor networks. $\Delta t_{\text{charge}} = 8$ seconds is the charging duration. Finally, \( n \) represents the path loss exponent, which quantifies how the signal weakens as the distance increases \cite{pathloss2013}.

\subsection{Time-Dependent System State Updates}

To provide explicit temporal modeling, we present the time-dependent update equations for energy and buffer dynamics during mobile agent operations.

\textbf{Energy Update Equations:}
The energy level of sensor node $i$ at time $t + \Delta t$ is updated according to:

\begin{equation}
\begin{split}
E_i(t + \Delta t) = \; & E_i(t) + P_{\text{effective}}(d_i, N_{\text{active}}) \\
                        & \times \Delta t_{\text{charge}} \times I_{\text{charge}}(t) \\
                        & - E_{\text{consumption}}(t)
\end{split}
\label{eq:energy_update}
\end{equation}
where $I_{\text{charge}}(t)$ is an indicator function equal to 1 if the mobile agent performs a charging action at time $t$, and 0 otherwise. The term $E_{\text{consumption}}(t)$ represents energy consumed by the sensor for transmission and reception activities during the time interval. $P_{\text{effective}}(d_i, N_{\text{active}})$ is defined in Equation~\eqref{eq:effective_power}, $d_i$ is the distance from the mobile agent to node $i$, and $N_{\text{active}}$ is the number of nodes being serviced simultaneously.

\textbf{Buffer Update Equations:}
The buffer occupancy of sensor node $i$ at time $t + \Delta t$ is updated according to:

\begin{equation}
\begin{split}
B_i(t + \Delta t) = \; & B_i(t) + D_{\text{generated}}(t) \\
                        & - R_{\text{node}}(N_{\text{active}}) \\
                        & \times \Delta t_{\text{collect}} \times I_{\text{collect}}(t)
\end{split}
\label{eq:buffer_update}
\end{equation}
where $I_{\text{collect}}(t)$ is an indicator function equal to 1 if the mobile agent performs data collection at time $t$, $\Delta t_{\text{collect}}$ is the data-collection duration, $D_{\text{generated}}(t)$ represents data packets generated by the sensor during the time interval, and $R_{\text{node}}(N_{\text{active}})$ is the per-node collection rate defined in Equation~\eqref{eq:data_rate_per_node}.

\textbf{Mobile Agent Energy Update:}
The mobile agent's energy evolves according to:

\begin{equation}
\begin{split}
E_{\text{MA}}(t + \Delta t) = \; & E_{\text{MA}}(t) + h_{\text{unit}} \times N_{\text{h}}(t, t+\Delta t) \\
                           & - E_{\text{operation}} \\
                           & - E_{\text{move}} \times d_{\text{traveled}}
\end{split}
\label{eq:ma_energy_update}
\end{equation}
where $N_{\text{h}}(t, t+\Delta t) \sim \text{Poisson}(\lambda_{\text{h}} \Delta t)$ is the number of stochastic energy-harvesting arrivals occurring in the interval $(t, t+\Delta t]$, with rate $\lambda_{\text{h}}$, and $h_{\text{unit}}$ is the fixed energy quantum delivered per arrival. $E_{\text{MA}}(t)$ is the mobile agent's current energy, $E_{\text{operation}}$ is the energy consumed during service operations (0.8 J for data collection, 0.4 J for charging), $E_{\text{move}} = 0.1$ J/m is the movement energy cost per meter, and $d_{\text{traveled}}$ is the distance traveled by the agent.

\subsection{Mobile Agent Model} \label{sec:mobile_agent_model}
In this study, it is assumed that the mobile agent and all sensor nodes use a single-antenna architecture \cite{javan2024code}. Due to fundamental limitations of single-antenna systems, the mobile agent performs data collection and sensor charging tasks sequentially. When using a single antenna, attempting SWIPT causes severe electromagnetic interference between power and data signals that cannot be effectively separated \cite{baikova2017interference}. The signal separation required for simultaneous operation would require complex hardware that significantly increases cost and power consumption \cite{perera2018swipt}. Additionally, attempting to transmit energy and data simultaneously leads to signal collisions and degrades the reliability of both operations, especially in networks with many nodes. To avoid these problems, our approach requires the mobile agent to alternate between data collection and sensor charging tasks. Upon arriving at a grid location, the agent first assesses local conditions and then performs either charging or data collection—never both—before moving to the next location. This sequential operation ensures signal integrity, keeps the system simpler, and improves both energy efficiency and overall reliability.

\subsubsection{Service Capacity and Coverage Specifications}
To address the quantitative requirements for mobile agent service capability, we provide explicit specifications for concurrent coverage and operational parameters. The mobile agent can serve a maximum of $M_{\text{max}} = 5$ sensor nodes simultaneously within its service radius of $R_{\text{service}} = 10$ meters. This limitation ensures manageable power distribution and prevents system overload during energy transfer operations. Each subdomain spans $L_{\text{grid}} \times L_{\text{grid}} = 20 \times 20$ m², containing an average of $N_{\text{avg}} = 4$ sensor nodes uniformly distributed. The maximum number of nodes that can be present in a single subdomain is $N_{\text{max}} = 7$ nodes, ensuring the mobile agent can always service all nodes within a subdomain during a single visit.

The mobile agent operates under strict timing constraints with charging operations requiring a fixed duration of $\Delta t_{\text{charge}} = 8$ seconds per subdomain, data collection operations lasting $\Delta t_{\text{collect}} = 6$ seconds per subdomain, and movement between adjacent subdomains taking $t_{\text{move}} = 5$ seconds. The charging rate varies with distance and number of active nodes according to:

\begin{equation}
P_{\text{effective}}(d_i, N_{\text{active}}) = \frac{P_{\text{charge}} \times 
\eta \times (d_0/d_i)^n}{N_{\text{active}}}
\label{eq:effective_power}
\end{equation}
where $d_i$ is the distance to node $i$, $P_{\text{charge}} = 1.5~\mathrm{J/s}$ is the charging power at the reference distance $d_0 = 1~\mathrm{m}$, $\eta$ is the energy transfer efficiency, and the total power is shared among $N_{\text{active}}$ nodes being charged simultaneously. Similarly, the data collection rate adapts based on the number of active nodes:

\begin{equation}
R_{\text{node}}(N_{\text{active}}) = \frac{R_{\text{collect}}}{N_{\text{active}}} = \frac{80 \text{ kbps}}{N_{\text{active}}}
\label{eq:data_rate_per_node}
\end{equation}
where $R_{\text{collect}} = 80$ kbps is the data collection rate ensuring fair bandwidth allocation among communicating sensors.

The mobile agent starts from the base station with initial energy of \( E_{\text{MA,init}} \) and alternates between two main tasks: charging sensors and collecting data within the grid where it is currently stationed using RF. In charging mode, the mobile agent operates at a rate of 1.5 Joules per second at 1 meter distance, with an effective coverage radius of 10 meters. It can charge multiple sensors simultaneously within this radius for a duration of 8 seconds per subdomain, serving up to 5 nodes in a single charging operation. This constitutes a one-to-many energy transfer relationship. In data collection mode, the agent sends a request to all nodes in the current subdomain and operates at an aggregate collection rate of 80~kbps for 6~s per subdomain. The aggregate rate is equally shared among the $N_{\mathrm{active}}$ communicating nodes according to Equation~(\ref{eq:data_rate_per_node}), providing fair bandwidth allocation among the nodes served during each visit. At the system-model level, the communication process is abstracted over this fixed collection interval, and separate MAC-layer contention, collision-resolution, and per-node scheduling mechanisms are not explicitly modeled.

In this study, it is assumed that the mobile agent is equipped with an energy harvesting module that allows it to collect energy from the environment during its operation. Since the agent performs two energy-intensive tasks—alternating between data collection and sensor charging—efficient energy management is essential. The quantitative impact of these operations on the agent's energy is significant: each charging operation transfers up to 12 Joules to sensors (at 1.5 J/s for 8 seconds), while data collection consumes approximately 0.8 Joules per subdomain. To support sustained operation, the harvested energy is stored in rechargeable batteries embedded within the agent. While the mobile agent may return to the base station for recharging when necessary, it primarily relies on environmental sources to replenish its energy. Recent advancements in energy harvesting technologies have made these sources increasingly viable for wireless applications \cite{sharma2023EH}. However, since energy harvesting in real-world conditions is inherently uncertain \cite{liu2023power}, this study assumes that the timing of energy arrivals follows a stochastic process, representing the random nature of environmental energy availability.

\subsubsection{Time Domain Operation Model}

To clarify the temporal dynamics of mobile agent operations, we provide explicit time domain specifications for system evolution and resource allocation.

\textbf{Action Set and Decision Logic:}
At each subdomain visit, the mobile agent selects exactly one action from the set $A = \{a_{\text{charge}}, a_{\text{collect}}\}$. The agent cannot perform both actions during a single visit. If the agent could perform both actions at every location, there would be no decision-making problem to solve - it would simply visit every grid and perform both operations, eliminating the need for intelligent resource allocation and learning-based optimization. Moreover, performing both operations at every location would waste significant energy and time resources, as many locations may require only one type of service based on local sensor conditions.

\textbf{Service Time Parameters:}
Each operation consumes fixed time regardless of the number of nodes served within the subdomain:
\begin{itemize}
    \item Charging operation: $\Delta t_{\text{charge}} = 8$ seconds per subdomain visit
    \item Data collection operation: $\Delta t_{\text{collect}} = 6$ seconds per subdomain visit  
    \item Movement time: $t_{\text{move}} = 5$ seconds between adjacent subdomains
\end{itemize}

These fixed durations ensure predictable network coverage cycles and prevent the agent from spending excessive time at high-demand locations while neglecting other network areas.

\textbf{Coverage Model:}
When the mobile agent performs an action in a subdomain, it affects all sensor nodes within that grid simultaneously through broadcast-based RF transmission. For charging operations, all nodes within the service radius ($R_{\text{service}} = 10$ meters) receive energy transfer simultaneously. For data collection, the agent requests and receives data from all nodes within range during the same operation period. This is not a sequential per-node process but rather a coverage-based simultaneous effect on all nodes present in the subdomain.

\textbf{System Time Evolution:}
The total time for one complete decision cycle is:

\begin{equation}
\begin{split}
T_{\text{cycle}} &= t_{\text{move}} + \max(\Delta t_{\text{charge}}, \Delta t_{\text{collect}}) \\
           &= 5 + 8 = 13 \text{ seconds}
\end{split}
\label{eq:tcycle}
\end{equation}

All time costs are explicitly counted in system evolution with no temporal simplifications. The simulation advances in discrete time steps accounting for movement, service duration, and decision-making phases.

\subsection{Sensor Node Model}

In this study, to reduce energy consumption, sensors do not immediately send their data to the base station after it is generated. Instead, they temporarily store the data in their buffer. It is assumed that sensors delay transmission until their buffer reaches a certain threshold, hoping that the mobile agent will arrive in time to collect the data. This way, the data can be delivered to the base station through the mobile agent, without relying on multi-hop communication, which consumes more energy across the entire network.

\begin{figure}[h!]
    \centering
    \includegraphics[width=0.50\textwidth]{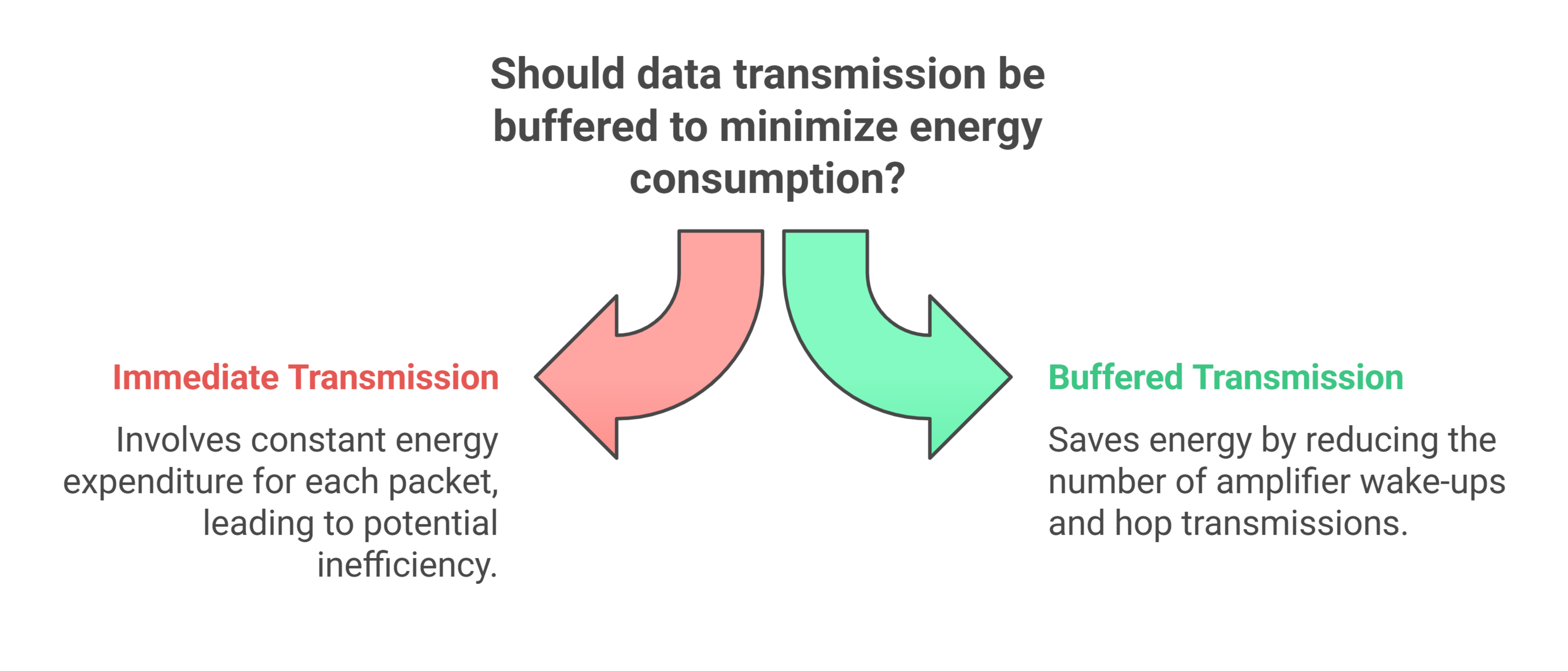}
    \caption{Comparison of immediate versus buffered transmission strategies. Immediate transmission incurs constant energy cost per packet due to repeated amplifier wake-ups, while buffered transmission reduces energy consumption by batching packets and minimizing multi-hop forwarding overhead.}
    \label{fig:buffer_model}
\end{figure}

The assumption of delaying data transmission until the buffer reaches a certain threshold is made for two main reasons. First, waking up the amplifier for each individual packet consumes a fixed amount of start-up energy \( E_0 \), making per-packet transmission inefficient \cite{article_amp}. By buffering multiple packets and transmitting them in a batch, energy usage can be significantly reduced. Second, multi-hop data transmission increases energy consumption, especially for nodes near the base station that must forward data from distant nodes. By waiting for the mobile agent to arrive and transmit data directly, the need for energy-intensive multi-hop routing is minimized—unless the buffer becomes full before the agent reaches the node \cite{Mehmood}. See Figure~\ref{fig:buffer_model}.

\subsection{Scalability Considerations} \label{sec:scalability_considerations}
This proposed approach is meant for one mobile agent in operation in a grid network. The problem with this single-agent architecture is that it works well only on a smaller or moderate scale; otherwise, it is not scalable at a large scale. The primary limiting factor of this approach is that when the network size is large, the distance between subdomains increases and thus might lead to buffer overflow in unvisited nodes and energy depletion in some important parts before the mobile agent visits them. In a 5 × 5 grid with 100–500 nodes, as tested in this study, one mobile agent will be sufficient to cover the whole network.

Expanding the proposed DCR-HRL approach to multi-agent systems could eliminate these shortcomings by having multiple agents that can concurrently explore different network areas. Potential benefits of such an expansion would include more rapid network exploration, lower service delay for each node, proper load balancing within the network, and increased robustness against failures of any agent. Nevertheless, multi-agent configurations with our hierarchical architecture also present a number of important issues. First, \textbf{coordination overhead} is introduced due to the need for multiple agents to exchange knowledge about the global network state to prevent unnecessary repeated visits to the same sub-domains. Second, \textbf{task conflicts} may arise when several agents simultaneously try to charge or gather data from the same node, wasting energy and reducing overall efficiency. Third, \textbf{load imbalance} may occur because agents pursue their own Q-learning strategies independently, thus over-serving some regions while under-serving others. Finally, \textbf{communication overhead} between agents is additional complexity and extra energy consumption, which is contrary to the energy-efficient goals of the framework. These scalability and coordination challenges motivate future extensions of DCR-HRL toward cooperative multi-agent reinforcement learning, as discussed in Section~\ref{sec:conclusion}.

\section{Analytical Models}\label{sec6}

In wireless sensor networks, the sensors that are closer to the base station usually run out of energy faster because they handle more data forwarding. If this imbalance is not managed well, it can cause network congestion, reduce coverage, and lead to early failure of the network. To address this problem, we use a smart and adaptive mobile agent that can sequentially collect data from busy areas and recharge sensors with low energy. This helps balance the energy usage across the network and improves its lifetime. To enable this behavior, the agent’s decision-making process is modeled as two sub-problems using the MDP, and a hierarchical Q-learning algorithm is applied to solve them at both the service-selection and movement-planning levels.

The rest of this section explains the analytical model in more detail. Section V-A presents the MDP formulation, Section V-B describes the system dynamics, Section V-C introduces the learning model, and Section V-D presents the mobile agent operation algorithm.

\subsection{Markov Decision Process Model}

In general, an MDP offers a formal framework for modeling sequential decision-making problems in environments where outcomes are partly governed by chance and partly influenced by the decision maker. In a discrete-time MDP, at each time step, the decision maker selects an action that affects the evolution of the stochastic system—meaning that both the next state and the reward depend on the chosen action. A hierarchical Markov Decision Process (HMDP) extends the classical MDP framework by organizing the decision-making process into multiple levels or phases, each responsible for a different type of action or abstraction.

Formally, an HMDP is structured as a set of interconnected sub-MDPs, where each subproblem is defined by a tuple $(S^i, A^i, P^i, R^i, \gamma)$, with $S^i$ representing the state space of subtask $i$, $A^i$ the corresponding action space, $P^i(s'|s,a)$ the transition probability, $R^i(s,a)$ the reward function, and $\gamma$ the shared discount factor. The primary objective remains to determine a policy that maximizes the cumulative reward, but this is achieved through the coordination of sub-policies at each level. Solving the HMDP involves identifying the optimal action sequences within each sub-MDP while accounting for how decisions at one level affect transitions and outcomes in another. Throughout this paper, subscript 1 refers to the lower layer (service selection), so $s_1$, $A_1$, $Q_1$, and $R_1$ correspond to charging and data collection decisions. Subscript 2 refers to the upper layer (movement planning), so $s_2$, $A_2$, $Q_2$, and $R_2$ correspond to movement decisions.

We model our problem using an HMDP, where the goal is to develop a strategy that helps the mobile agent choose the best action at each phase and maximize the total reward. Although the theoretical MDP framework includes transition probabilities $P^i(s'|s,a)$, our Q-learning-based solution is model-free and learns optimal policies through direct environment interaction without requiring explicit probability models—a key advantage in WSN scenarios where node failures and energy dynamics are difficult to model analytically. In this framework, the decision process is split into two interconnected levels: one for deciding whether to charge sensors or collect their data, and another for determining where the agent should move next. The mobile agent (as the decision maker) learns to coordinate these service and movement decisions to operate efficiently across the network. A detailed explanation of the system formulation is provided in the following sections.

\subsubsection{Decision Epochs}

A decision epoch is a specific moment when the mobile agent is allowed to make a decision that may change the system’s current state. In our problem, the agent alternates between two decision epochs: at the upper layer, it selects its next destination to move to; and at the lower layer, once it stops at that destination, it decides whether to charge the nearby nodes or collect their data. Each of these decisions belongs to a different phase in the hierarchical decision-making process.

\subsubsection{State Space}
To ensure efficient learning and avoid the curse of dimensionality, all state variables are normalized to the range [0,1].

\vspace{1em}
\noindent\textbf{a: Upper Layer State Space (Normalized)}

The upper layer state representation captures the mobile agent's spatial context and the energy-buffer status of neighboring grids, as defined in Equation~\ref{eq:upper_state}:

\begin{equation}
\begin{split}
s_2 = \Bigg( & \frac{x_{\text{grid}}}{W}, \frac{y_{\text{grid}}}{H}, 
               \left\{\frac{E_{\text{neighbor},i}}{E_{\text{max}}}\right\}_{i \in \mathcal{N}}, \\
             & \left\{\frac{B_{\text{neighbor},i}}{B_{\text{max}}}\right\}_{i \in \mathcal{N}}, 
               \frac{E_{\text{agent}}}{E_{\text{MA},\text{init}}} \Bigg)
\end{split}
\label{eq:upper_state}
\end{equation}

where $W$ and $H$ are the grid dimensions, $x_{\text{grid}}$ and $y_{\text{grid}}$ are the agent's current grid coordinates, $\mathcal{N}$ is the set of neighboring grids, $E_{\text{max}}$ is the maximum sensor energy capacity, $B_{\text{max}}$ is the maximum buffer size, $B_{\text{neighbor},i}$ denotes the buffer occupancy in the neighboring grid, $E_{\text{neighbor},i}$ is the energy occupancy of the neighboring grid, and $E_{\text{MA},\text{init}}$ is the agent's initial energy.

The framework does not rely on offline, predefined network knowledge fixed in advance. Instead, $E_{\text{neighbor},i}$ and $B_{\text{neighbor},i}$ for all $i \in \mathcal{N}$ are obtained through lightweight status reporting: each sensor appends a 32-bit field carrying its subdomain index together with its quantized energy and buffer occupancy to the data packets it is already forwarding. Sensors in the neighboring subdomains deliver these fields to the mobile agent over the existing multi-hop fabric, and the agent averages the reports received. These indicators are thus refreshed as a by-product of normal network operation, without any dedicated control transmissions. At the node communication range $R_{\text{comm}} = 15$~m, a report adds $2.32\times10^{-6}$~J. Relative to a single 4000-bit data packet, this is exactly 0.8\%, independently of the link distance, so the overhead is negligible with respect to the sensors' energy budget. Destination selection is performed on online information acquired during operation, not on a precomputed itinerary or a prior map of the network. The only knowledge carried across decision epochs is the learned Q-tables themselves, which encode the long-term value of each movement; because these values are learned rather than hand-coded, the policy responds to the state observed at each decision epoch instead of executing a fixed schedule.

\vspace{1em}
\noindent\textbf{b: Lower Layer State Space (Normalized)}

After the mobile agent arrives at a grid location, the lower layer state aggregates local conditions to determine which service action to perform, as defined in Equation~\ref{eq:lower_state}:

\begin{equation}\label{eq:lower_state}
s_1 = \left(\frac{\text{Grid\_Index}}{\text{Total\_Grids}}, \frac{E_{\text{avg}}}{E_{\text{max}}}, \frac{B_{\text{avg}}}{B_{\text{max}}}\right)
\end{equation}
where $\text{Grid\_Index}$ is the index of the agent's current grid, $\text{Total\_Grids}$ is the total number of grids (25 in this study), $E_{\text{avg}}$ is the average energy level of sensors in the current grid, and $B_{\text{avg}}$ is the average buffer occupancy of sensors in the current grid. This normalization ensures that the Q-learning algorithm explores the state space efficiently and converges faster.

\subsubsection{Action Space}
The action space represents the set of possible decisions the mobile agent can make in each state.

\vspace{1em}
\noindent\textbf{a: Upper Layer Action Space}\\

The action space at the upper layer consists of the four main directions in which the mobile agent can move from its current position, as defined in Equation~\eqref{eq:action_space_phase2}.
\begin{equation}
A_2 = \{a_{\text{move}}^i \mid i \in \{\text{up}, \text{down}, \text{left}, \text{right}\}\}
\label{eq:action_space_phase2}
\end{equation}

\vspace{1em}
\noindent\textbf{b: Lower Layer Action Space}\\
At each grid location, the mobile agent must choose exactly ONE action from the following \textbf{mutually exclusive} options:
\begin{equation}
A_1 = \{a_{\text{charge}}, a_{\text{collect}}\}
\label{eq:action_space_phase1}
\end{equation}
These actions are mutually exclusive—the agent cannot perform both simultaneously due to single-antenna constraints. The agent makes this decision based on current network conditions and learned Q-values.

\textbf{1. Charging Action ($a_{\text{charge}}$):}  
When this action is selected, the mobile agent charges sensors within the grid for a fixed duration of 8 seconds. It monitors their energy levels and gives priority to those with the lowest remaining energy. The amount of energy transferred per sensor is determined by Equation~\eqref{eq:energy_charged_}.
\begin{equation}
\label{eq:energy_charged_}
E_{\text{charged}} = \eta \cdot 
\min(E_{\text{max}} - E_{\text{sensor}}, E_{\text{agent}})
\end{equation}
where $\eta$ represents the energy transfer efficiency (0.5), $E_{\text{max}}$ denotes the maximum energy capacity of the sensor, $E_{\text{sensor}}$ indicates the current energy level of the sensor, $E_{\text{agent}}$ corresponds to the remaining energy of the mobile agent. The charging visit itself has a fixed duration of $\Delta t_{\mathrm{charge}} = 8$~seconds.

\textbf{2. Data Collection Action ($a_{\text{collect}}$):}  
When this action is selected, the mobile agent collects data from sensors in the grid for a fixed duration of 6 seconds. It assesses the buffer levels of sensors and gives priority to those with higher occupancy. The amount of data collected is calculated in Equation~\eqref{eq:data_collected}.
\begin{equation}
\label{eq:data_collected}
D_{\text{collected}} = \min(B_{\text{sensor}}, B_{\text{agent}}, R_{\text{collect}} \cdot \Delta t_{\text{collect}})
\end{equation}
where $B_{\text{sensor}}$ represents the current buffer occupancy of the sensor, $B_{\text{agent}}$ denotes the maximum data capacity the mobile agent can handle, which is assumed unbounded in this study, $R_{\text{collect}} = 80$ kbps is the data collection rate, and $\Delta t_{\text{collect}} = 6$ seconds is the collection duration.

\textbf{Action Selection:} The agent selects one action at each grid location based on the Q-learning policy~\cite{malekijou2023q}. After completing the selected action (either charging OR collection), the agent proceeds to the movement phase to select its next destination. This sequential, single-action-per-location approach ensures no interference between charging and data signals while maintaining system efficiency.

\subsubsection{Reward function}

The reward function is designed to encourage efficient behavior while discouraging inefficient actions. 

\vspace{1em}
\noindent\textbf{a: Upper Layer Reward Function}\\

The upper layer reward function is defined in Equation~\eqref{eq:phase2_reward}, which combines charging potential, data collection potential, and movement cost:

\begin{equation}
\begin{split}
R_2(s_2, a_{\text{move}}^i) = \; & R_{\text{charge}}(s_2, a_{\text{move}}^i) \\
                           & + R_{\text{collect}}(s_2, a_{\text{move}}^i) \\
                           & - R_{\text{cost}}(s_2, a_{\text{move}}^i)
\end{split}
\label{eq:phase2_reward}
\end{equation}

where $R_2(s_2, a_{\text{move}}^i)$ denotes the overall reward for a movement action in state $s_2$. In this formula, $R_{\text{charge}}(s_2, a_{\text{move}}^i)$ represents the charging potential reward, $R_{\text{collect}}(s_2, a_{\text{move}}^i)$ the potential data collection reward, and $R_{\text{cost}}(s_2, a_{\text{move}}^i)$ the movement cost.

The charging potential reward is explained in Equation~\eqref{eq:charging_reward}, encouraging movement to low-energy grids:

\begin{equation}
\begin{split}
R_{\text{charge}}(s_2, a_{\text{move}}^i) = \; & w_{\text{charge}} \cdot \eta \\
& \cdot \min(E_{\text{max}} - E_{\text{neighbor},i},\, E_{\text{agent}}) \\
& \cdot \frac{E_{\text{threshold}} - E_{\text{neighbor},i}}{E_{\text{threshold}}} \\
& \cdot I(E_{\text{neighbor},i} < E_{\text{threshold}})
\end{split}
\label{eq:charging_reward}
\end{equation}

where $w_{\text{charge}}$ indicates the weight assigned to charging, $E_{\text{threshold}}$ is the critical low-energy level for charging priority, $\eta$ shows the energy transfer efficiency, and $\min(E_{\text{max}} - E_{\text{neighbor},i}, E_{\text{agent}})$ reflects the maximum energy that can be delivered, limited by both the sensor's capacity $(E_{\text{max}} - E_{\text{neighbor},i})$ and the mobile agent's available energy $E_{\text{agent}}$. The term $\frac{E_{\text{threshold}} - E_{\text{neighbor},i}}{E_{\text{threshold}}}$ normalizes the energy deficit relative to the threshold $E_{\text{threshold}}$, while $I(E_{\text{neighbor},i} < E_{\text{threshold}})$ is an indicator function that equals 1 when the sensor energy $E_{\text{neighbor},i}$ is below $E_{\text{threshold}}$ (and 0 otherwise).

We integrate the stochastic harvesting process $N_{\text{h}}(t, t+\Delta t)$ (Eq.~\eqref{eq:ma_energy_update}) into the framework via state augmentation; its cumulative effect is captured through the state variable $E_{\text{agent}}/E_{\text{MA},\text{init}}$ in $s_2$ (Eq.~\eqref{eq:upper_state}), which reflects the agent's true energy level after all harvesting arrivals and consumption up to the current decision epoch. This constitutes a state-augmentation treatment of the stochastic harvesting process, consistent with the model-free nature of our Q-learning formulation: the agent need not know the underlying arrival rate $\lambda_{\text{h}}$ to act optimally, only the resulting energy state, which directly determines the charging-related reward terms in $R_{\text{charge}}(s_2, a_{\text{move}}^i)$ (Eq.~\eqref{eq:charging_reward}) through the term $\min(E_{\text{max}} - E_{\text{neighbor},i}, E_{\text{agent}})$. In this way, the reward function inherits the stochasticity of the harvesting process indirectly but completely through $E_{\text{agent}}$, rather than through an explicit dependence on $\lambda_{\text{h}}$.

The data collection reward, defined in Equation~\eqref{eq:collection_reward}, favors movement toward grids with high buffer levels:

\begin{equation}
\begin{split}
R_{\text{collect}}(s_2, a_{\text{move}}^i) = \; & w_{\text{collect}} \cdot B_{\text{neighbor},i} \\
& \cdot \frac{B_{\text{neighbor},i} - B_{\text{threshold}}}{B_{\text{max}} - B_{\text{threshold}}} \\
& \cdot I(B_{\text{neighbor},i} > B_{\text{threshold}})
\end{split}
\label{eq:collection_reward}
\end{equation}

where $w_{\text{collect}}$ indicates the weight given to data collection, $B_{\text{threshold}}$ is the critical high-buffer level for collection priority, and $\frac{B_{\text{neighbor},i} - B_{\text{threshold}}}{B_{\text{max}} - B_{\text{threshold}}}$ normalizes the buffer level above the threshold $B_{\text{threshold}}$ relative to the maximum buffer capacity $B_{\text{max}}$. The indicator function $I(B_{\text{neighbor},i} > B_{\text{threshold}})$ equals 1 when the buffer occupancy exceeds $B_{\text{threshold}}$.

Finally, the movement cost component, defined in Equation~\eqref{eq:movement_cost}, penalizes the distance traveled:

\begin{equation}
\begin{split}
R_{\text{cost}}(s_2, a_{\text{move}}^i) = \; & \lambda_{\text{move}} \\
& \cdot \, d\big((x_{\text{grid}}, y_{\text{grid}}), \\
& \quad (x_{\text{grid}}, y_{\text{grid}}) + \Delta_i \big)
\end{split}
\label{eq:movement_cost}
\end{equation}

In this expression, $\lambda_{\text{move}}$ indicates the penalty coefficient for movement, and $d\big((x_{\text{grid}}, y_{\text{grid}}), (x_{\text{grid}}, y_{\text{grid}}) + \Delta_i\big)$ shows the distance traveled from the current grid $(x_{\text{grid}}, y_{\text{grid}})$ to the neighboring grid determined by the offset $\Delta_i$.

\vspace{1em}
\noindent\textbf{b: Lower Layer Reward Function}\\

The total reward $r$ at the lower layer is formulated in Equation~\eqref{eq:reward_first}.

\begin{equation}
\label{eq:reward_first}
    r = r_{\text{action}} + r_{\text{efficiency}} + r_{\text{penalty}}
\end{equation}

The components of the reward function are defined as follows:

\begin{itemize}
    \item \textbf{Action Reward} ($r_{\text{action}}$) is given by Equation~\eqref{eq:areward}.
    \begin{equation}
    \label{eq:areward}
    r_{\text{action}} =
    \begin{cases}
    \lambda_{\text{charge}} \times E_{\text{charged}}, & \text{if } a = \text{Charge} \\
    \lambda_{\text{collect}} \times D_{\text{collected}}, & \text{if } a = \text{Collect} \\
    \end{cases}
    \end{equation}

    \item \textbf{Efficiency Reward} ($r_{\text{efficiency}}$) is defined by Equation~\eqref{eq:ereward}, considering charging near the base station and data collection far from the base.
    \begin{equation}
    \label{eq:ereward}
    r_{\text{efficiency}} =
    \begin{cases}
    \lambda_{\text{near}} \times E_{\text{charged}},  & \text{if } a = \text{Charge} \\ 
    \lambda_{\text{far}} \times D_{\text{collected}}, & \text{if } a = \text{Collect}
    \end{cases}
    \end{equation}

    \item \textbf{Penalty} ($r_{\text{penalty}}$) is computed by Equation~\eqref{eq:pareward}, considering data collection near the base station.
    \begin{equation}
    \label{eq:pareward}
    r_{\text{penalty}} =
    \lambda_{\text{penalty}} \times D_{\text{collected}},
    \end{equation}
\end{itemize}
where $a$ represents the action taken, $E_{\text{charged}}$ indicates the amount of energy transferred to sensor nodes, $D_{\text{collected}}$ denotes the amount of data collected from nodes, and $\lambda_{\text{charge}}$, $\lambda_{\text{collect}}$, $\lambda_{\text{near}}$, and $\lambda_{\text{far}}$ are positive weighting constants, while $\lambda_{\text{penalty}}$ is a negative weighting constant that balances the contribution of each reward component.

The terms "near the base" and "far from the base" are defined based on a distance threshold $d_{\text{thresh}}$. A grid cell is considered "near the base" if it is within $d_{\text{thresh}}$ units of the base station; otherwise, it is classified as "far from the base."

\vspace{1em}
\noindent\textbf{c: Inter-Layer Reward Coordination}\\

To ensure coherent behavior across both levels, we introduce a global reward function \(G\) that combines the rewards from both levels:

\begin{equation}
G(s_1, a_1, s_2, a_2) = 
\beta_1 \cdot R_1(s_1, a_1) + \beta_2 \cdot R_2(s_2, a_2)
\label{eq:global_reward}
\end{equation}
where \(\beta_1\) and \(\beta_2\) indicate weighting factors that balance the importance of service and movement decisions, respectively,  
and \(\beta_1 + \beta_2 = 1\) ensures proper normalization.

The weighting factors are dynamically adjusted based on network conditions described in Equations~\eqref{eq:beta1} and \eqref{eq:beta2}:

\begin{equation}
\begin{split}
\beta_1 = \; & \sigma\Big( w_1 \cdot \frac{E_{\text{threshold}} - E_{\text{avg}}}{E_{\text{threshold}}} \\
             & + \, w_2 \cdot \frac{B_{\text{avg}} - B_{\text{threshold}}}{B_{\text{max}}} \Big)
\end{split}
\label{eq:beta1}
\end{equation}

\begin{equation}
\beta_2 = 1 - \beta_1
\label{eq:beta2}
\end{equation}

where $\sigma$ indicates the sigmoid function ensuring the output remains in the interval $[0,1]$,  
$w_1$ and $w_2$ indicate the relative importance of energy and buffer states, respectively.  
When energy levels are critical $(E_{\text{avg}} \ll E_{\text{threshold}})$, $\beta_1$ increases to prioritize service decisions.  
When buffer levels are high $(B_{\text{avg}} \gg B_{\text{threshold}})$, $\beta_1$ also increases to prioritize data collection.  
Otherwise, $\beta_2$ increases, granting more importance to movement actions.

This dynamic reward balancing ensures that the mobile agent adapts its behavior to current network conditions, maintaining coordinated decision-making between service and movement phases.

\subsection{System Dynamics}
While Q-learning does not require explicit knowledge of transition probabilities, we briefly describe the system dynamics for completeness. The mobile agent learns these dynamics implicitly through interaction with the environment.

\subsubsection{State Evolution}
The system state evolves based on:
\begin{itemize}
    \item \textbf{Energy dynamics:} Node energy decreases due to transmission and increases when charged by the agent
    \item \textbf{Buffer dynamics:} Buffers fill as sensors generate data and empty when the agent collects data
    \item \textbf{Time dynamics:} Each action (charging or collection) requires specific time duration and must be completed sequentially
\end{itemize}

The Q-learning algorithm discovers the optimal policy by observing state transitions and rewards, without needing an explicit model of these dynamics. This model-free approach is particularly suitable for WSNs where exact transition dynamics are difficult to model due to environmental uncertainties and node failures.

\subsubsection{Optimal Policy}

In the context of our hierarchical MDP formulation, we define a composite policy. \(\pi\) is defined in Equation~\eqref{eq:composite_policy}:

\begin{equation}
\pi = (\pi_1, \pi_2)
\label{eq:composite_policy}
\end{equation}
where $\pi_1 : S_1 \rightarrow A_1$ indicates the service decision policy, and $\pi_2 : S_2 \rightarrow A_2$ indicates the movement decision policy.

The objective is to determine the optimal composite policy \(\pi^* = (\pi_1^*, \pi_2^*)\) that maximizes the expected cumulative global reward described in Equation~\eqref{eq:optimal_policy_objective}.

\begin{equation}
\begin{aligned}
\pi^* = \arg\max_{\pi} \mathbb{E} \Big[ \sum_{t} \gamma^t \cdot \\
\hspace{2em} G(s_1^t, a_1^t, s_2^t, a_2^t) \Big]
\end{aligned}
\label{eq:optimal_policy_objective}
\end{equation}
where \(G(s_1, a_1, s_2, a_2)\) indicates the global reward function defined in Equation~\eqref{eq:global_reward}, which combines phase-specific rewards using dynamic weighting factors \(\beta_1\) and \(\beta_2\).

Due to the alternating nature of decision epochs between phases, the optimization problem is inherently coupled. The optimal policy in Phase 1 depends on the value function of Phase 2, and vice versa, as shown in Equation~\eqref{eq:pi1_opt}.

\begin{equation}
\begin{aligned}
\pi_1^*(s_1) = \arg\max_{a_1 \in A_1} \Big[ R_1(s_1, a_1) + 
\gamma \cdot \max_{a_2 \in A_2} Q_2(s_2, a_2) \Big]
\end{aligned}
\label{eq:pi1_opt}
\end{equation}

\begin{equation}
\begin{aligned}
\pi_2^*(s_2) = \arg\max_{a_2 \in A_2} \Big[ R_2(s_2, a_2) + 
\gamma \cdot \max_{a_1 \in A_1} Q_1(s_1', a_1) \Big]
\end{aligned}
\label{eq:pi2_opt}
\end{equation}

This hierarchical policy structure ensures that service and movement decisions are jointly optimized. The agent dynamically adapts its decision-making behavior to network conditions through the reward weights \(\beta_1\) and \(\beta_2\), effectively balancing energy provisioning and data collection across the network.

\subsubsection{Inter-Layer Coupling: Formal Analysis}
The coupling between the two layers follows directly from the definition of the value functions. Let $V_1(s_1) = \max_{a_1} Q_1(s_1,a_1)$ and $V_2(s_2) = \max_{a_2} Q_2(s_2,a_2)$ denote the value of the service-selection and movement-planning layers,
respectively. By the Bellman optimality condition, the action-value function of each layer is defined in terms of the \emph{other} layer's value function:
\begin{equation}
Q_1(s_1, a_1) = R_1(s_1, a_1) + \gamma \, V_2\big(M_{12}(s_1, a_1)\big)
\label{eq:q1_bellman_derivation}
\end{equation}
\begin{equation}
Q_2(s_2, a_2) = R_2(s_2, a_2) + \gamma \, V_1\big(M_{21}(s_2, a_2)\big)
\label{eq:q2_bellman_derivation}
\end{equation}
where $M_{12}$ and $M_{21}$ are the state-mapping functions that carry the outcome of a service action into the subsequent movement state and the outcome of a movement action into the subsequent service state. Because $V_2$ appears inside the definition of $Q_1$ and $V_1$ appears inside the definition of $Q_2$, the two value functions are mutually recursive. Their consequences for action selection are immediate: taking the $\arg\max$ over $a_1$ shows that the service-selection layer does not choose the action with the largest immediate reward $R_1$, but the action whose resulting state offers the movement layer the greatest subsequent value, and symmetrically for the movement layer. A change in either layer's learned values therefore alters the other layer's greedy action at the next decision epoch. The incremental Q-learning updates in Equations~\eqref{eq:q1_update} and \eqref{eq:q2_update} are the sample-based form of this coupling, learned through interaction.

The dynamic weights $\beta_1$ and $\beta_2$ serve a distinct purpose from this coupling and are deliberately kept outside it. The inter-layer influence described above is carried entirely by the bootstrapped value terms, and each layer is trained on its own reward signal; the weights instead define the global objective $G$ against which the hierarchical design is posed, expressing how much priority service provisioning versus movement planning should receive under the prevailing network condition. Their sigmoid form is chosen for two properties: $\beta_1$ increases monotonically as the energy deficit or buffer excess worsens, so the shift in priority is predictable, and the sigmoid is smooth, so the weights vary gradually near the energy and buffer thresholds instead of switching discontinuously. Keeping the weights separate from the bootstrapped terms preserves the convergence properties of the underlying tabular updates, which would otherwise be disturbed by a reward scaling that changes with the network state.

\subsection{Learning Model}
Several approaches, such as hierarchical policy optimization and value-based decomposition, are used to solve hierarchical MDPs. However, these model-based techniques  usually need a lot of information about how the system works. In many real-world applications, including our scenario, it is difficult to model the environment's behavior. For example, it is not clear with what probability the sensors will shut down or how likely they are to send data, which makes it challenging to derive an optimal policy using traditional methods. As an alternative, reinforcement learning (RL) offers a practical solution for such decision-making problems, where theoretical optimization is hard to achieve. 

The goal of RL is to find the best policy incrementally in terms of maximizing cumulative reward (\( G_t \)), which is calculated in Equation~\eqref{eq:cumulative_reward}.

\begin{equation}
G_t = \sum_{k=0}^\infty \gamma^k r_{t+k+1} 
\label{eq:cumulative_reward}
\end{equation}
where \( G_t \) represents the cumulative return at time step \( t \), \( \gamma \) is the discount factor (\( 0 \leq \gamma \leq 1 \)), which determines the weight of future rewards relative to immediate rewards, and \( r_{t+k+1} \) is the reward received at time step \( t+k+1 \). A higher value of \( \gamma \) prioritizes long-term rewards, while a lower value favors immediate rewards \cite{Yau,sutton2018}.

Among RL algorithms, Q-learning is a widely used method that enables an agent to learn an optimal policy without requiring any prior knowledge of transition dynamics or probabilities. In this work, the mobile agent maintains two separate Q-tables and updates their values according to the actions it performs, including charging sensor nodes and collecting data. Each action yields a reward, which is then assigned to the grid where the service was carried out. After completing a task, the agent moves to the next grid expected to offer the highest reward, as determined during the movement phase of the hierarchical MDP.

This hierarchical structure allows the agent to separate service-level decisions from movement-level decisions, resulting in more efficient coordination across tasks and locations. The Q-learning process is applied at both levels, where each phase maintains its own Q-function. However, the two levels are interconnected through phase transitions and mutual influence of Q-values. This learning process continues until the agent converges to an optimal policy. The ultimate objective is to maximize the cumulative reward, enabling the mobile agent to navigate efficiently, extend the network's lifetime, improve throughput, and balance energy consumption and data loads across the network.

Initially, the Q-tables are set to zeros or small random values, corresponding to two initialization methods (deterministic and stochastic), as shown in Equation~\eqref{eq:q_init}:
\begin{equation}
Q_{\text{init}}(s, a) = 
\begin{cases}
0, & \text{deterministic} \\
\text{small random value}, & \text{stochastic}
\end{cases}
\label{eq:q_init}
\end{equation}

As the mobile agent interacts with the environment, the Q-values are updated according to the Bellman equation described in Equation~\eqref{eq:bellman_general}:

\begin{equation}
\label{eq:bellman_general}
Q(s, a) \leftarrow Q(s, a) + \alpha \big[ r + \gamma \max_{a'} Q(s', a') - Q(s, a) \big]
\end{equation}

where $\alpha$ is the learning rate determining how much new information overrides old information, $r$ is the immediate reward received after taking action $a$ in state $s$, $\gamma$ is the discount factor balancing immediate and future rewards, and $\max_{a'} Q(s', a')$ represents the maximum expected future reward achievable in the next state $s'$.

Given our two-level MDP formulation, the Q-learning process must be adapted to handle the alternating decision sequence. The mobile agent maintains two separate Q-tables:

\begin{itemize}
\item \textbf{Upper level Q-table:} Each entry is indexed by a state-action pair, where the state includes the current grid position $x_{\text{grid}}, y_{\text{grid}}$, neighboring energy levels $\{E_{\text{neighbor}}\}$, and neighboring buffer occupancies $\{B_{\text{neighbor}}\}$. The actions represent movement decisions in the four cardinal directions: up, down, left, and right.
\item \textbf{Lower level Q-table:} Each entry is indexed by a state-action pair, where the state is defined by the agent's grid position $x_{\text{grid}}, y_{\text{grid}}$, the average energy level $E_{\text{avg}}$, and the average buffer occupancy $B_{\text{avg}}$. The associated actions correspond to service-level decisions such as charging or data collection.
\end{itemize}

The Q-values are updated based on the following rules: 

\begin{equation}
\begin{split}
Q_1(s_1, a_1) \leftarrow \; & Q_1(s_1, a_1) + \alpha \cdot \big[ R_1(s_1, a_1) \\
                             & + \gamma \cdot \max_{a_2 \in A_2} Q_2(s_2, a_2) \\
                             & - Q_1(s_1, a_1) \big]
\end{split}
\label{eq:q1_update}
\end{equation}

\begin{equation}
\begin{split}
Q_2(s_2, a_2) \leftarrow \; & Q_2(s_2, a_2) + \alpha \cdot \big[ R_2(s_2, a_2) \\
                             & + \gamma \cdot \max_{a_1 \in A_1} Q_1(s_1', a_1) \\
                             & - Q_2(s_2, a_2) \big]
\end{split}
\label{eq:q2_update}
\end{equation}

To balance exploration and exploitation in both decision phases, an $\epsilon$-greedy policy is employed as described in Equation~\eqref{eq:epsilon_greedy}:

\begin{equation}
a_t = 
\begin{cases}
\arg\max_a Q(s_t, a), & \text{with probability } 1 - \epsilon \\
\text{random action}, & \text{with probability } \epsilon
\end{cases}
\label{eq:epsilon_greedy}
\end{equation}

\textbf{Hyperparameter Selection:} We set $\gamma = 0.9$ to emphasize long-term network lifetime over immediate gains, as current decisions affect network states over hundreds of epochs. The learning rate $\alpha = 0.1$ provides stable convergence for our 13,750-entry Q-table while maintaining sufficient adaptability to dynamic network conditions.

The value of $\epsilon$ decreases over time according to Equation~\eqref{eq:epsilon_decay}:

\begin{equation}
\epsilon = \epsilon_{\text{min}} + (\epsilon_{\text{max}} - \epsilon_{\text{min}}) \cdot e^{-\delta t}
\label{eq:epsilon_decay}
\end{equation}
where $\epsilon_{\text{min}}$ is the minimum exploration rate, $\epsilon_{\text{max}}$ is the initial exploration rate, $\delta$ is the decay factor, and $t$ is the current time step.

As the mobile agent gains experience through interactions with the WSN environment, the Q-table converges toward the optimal policy, effectively learning the best decision-making strategy for charging nodes, collecting data, and navigating through the network to maximize overall performance and network lifetime.

\subsubsection{Q-table Specification and Algorithm Scalability} \label{sec:qtable_spec}
To ensure efficient learning and tractable computation, we implement a normalized and discretized state space for our Q-learning approach. All state variables are normalized to [0,1] and discretized into 5 bins each: [0-0.2], [0.2-0.4], [0.4-0.6], [0.6-0.8], [0.8-1.0] for energy and buffer levels, while grid positions remain inherently discrete (25 locations in a $5 \times 5$ grid) shown in Equations~\eqref{eq:norm_upper} and~\eqref{eq:norm_lower}.

\begin{equation}
\begin{split}
s_2^{\text{norm}} = \Bigg( & \frac{x_{\text{grid}}}{5}, \frac{y_{\text{grid}}}{5}, \frac{E_{\text{neighbor}}}{20}, \\
                    & \frac{B_{\text{neighbor}}}{100}, \frac{E_{\text{agent}}}{50} \Bigg)
\end{split}
\label{eq:norm_upper}
\end{equation}

\begin{equation}
s_1^{\text{norm}} = \left(\frac{\text{Grid Index}}{25}, \frac{E_{\text{avg}}}{20}, \frac{B_{\text{avg}}}{100}\right)
\label{eq:norm_lower}
\end{equation}

The hierarchical structure employs two Q-tables: an upper layer with 12,500 entries (25 grids $\times$ 5 energy bins $\times$ 5 buffer bins $\times$ 5 agent energy bins $\times$ 4 movement actions) and a lower layer with 1,250 entries (25 grids $\times$ 5 energy bins $\times$ 5 buffer bins $\times$ 2 service actions), totaling 13,750 entries and requiring 110~KB of memory. Training uses learning rate $\alpha = 0.1$, discount factor $\gamma = 0.9$, and $\epsilon$-greedy exploration decaying from 0.3 to 0.01.

The state space size of 13,750 entries remains tractable for tabular Q-learning, allowing exact value function representation without approximation errors. For larger-scale deployments, Deep Q-Networks (DQN) or Actor-Critic methods would be more suitable to handle continuous state spaces and scale to larger networks. However, for our network size, tabular methods offer guaranteed convergence to the optimal policy and interpretable Q-values for analysis. The training process uses episodes of 500 decision epochs each, with a total of 1000 training episodes. The convergence criterion is defined as $|\Delta Q_{\text{avg}}| < 0.01$ maintained for 50 consecutive episodes, where $\Delta Q_{\text{avg}}$ represents the average absolute change in Q-values across all state-action pairs between consecutive episodes. This criterion ensures policy stability before terminating the training process.

\subsection{Mobile Agent Operation Algorithm}

In this section, we explain how the mobile agent works under the proposed hierarchical decision-making framework. To show this behavior clearly, we present an algorithm that guides the agent through its two main tasks: choosing what service to provide and deciding where to move next. The agent switches between these tasks in an organized way, making decisions based on the network’s real-time needs.

After completing a service in its current location, the mobile agent needs to decide where to go next. It examines the four neighboring grids—up, down, left, and right—and estimates how helpful each one would be in terms of future services. For each grid, the agent considers how much energy it could recharge or how much data it could collect. These estimates are combined into a value that reflects the grid’s usefulness. The agent then uses a learned Q-function to choose the grid with the highest value, helping it move in a smart and efficient way that improves overall network performance.

When the mobile agent stops in a grid, it decides which service to perform based on the needs of the nearby sensors. If sensors are low on energy, the agent chooses to charge them to avoid failures. If the data buffers are nearly full, the agent collects data to reduce congestion. These decisions are made using a reward-based model that encourages the agent to provide the right service at the right time. This flexible and adaptive behavior allows the agent to respond to changing conditions and keep the network running smoothly and efficiently.

Algorithm~\ref{alg:HierarchicalQLearning} summarizes the overall decision-making procedure of the mobile agent, encompassing both routing (upper layer) and service provisioning (lower layer) within the proposed hierarchical Q-learning framework.

\begin{algorithm}[h]
\caption{Hierarchical Q-Learning for Mobile Agent}
\label{alg:HierarchicalQLearning}
\begin{algorithmic}[1]
\Require Initialize \( Q_1(s_1, a_1) \), \( Q_2(s_2, a_2) \), learning rate \( \alpha \), discount factor \( \gamma \), exploration rates \( \epsilon_1, \epsilon_2 \)
\For{each episode}
    \State Initialize agent position \( (x_{\text{grid}}, y_{\text{grid}}) \), energy and buffer states
    \State Construct initial Phase 1 state \( s_1 = (x_{\text{grid}}, y_{\text{grid}}, E_{\text{avg}}, B_{\text{avg}}) \)
    \While{episode not terminated}
        \State \textbf{// Phase 1: Service Decision}
        \State Select action \( a_1 \in \{\text{Charge}, \text{Collect}\} \) using \( \epsilon_1 \)-greedy over \( Q_1(s_1, a_1) \)
        \State Execute \( a_1 \), update sensor states, observe reward \( r_1 \)
        \State Map to next movement state: \( s_2 = M_{12}(s_1, a_1) \)
        \State \( Q_1(s_1, a_1) \gets Q_1(s_1, a_1) + \alpha \cdot \big[ r_1 + \gamma \cdot \max_{a_2} Q_2(s_2, a_2) - Q_1(s_1, a_1) \big] \)
        \State \textbf{// Phase 2: Movement Decision}
        \State Select direction \( a_2 \in \{\text{Up}, \text{Down}, \text{Left}, \text{Right}\} \) using \( \epsilon_2 \)-greedy over \( Q_2(s_2, a_2) \)
        \State Move agent to new grid, update energy level, observe reward \( r_2 \)
        \State Map to service state: \( s_1' = M_{21}(s_2, a_2) \)
        \State \( Q_2(s_2, a_2) \gets Q_2(s_2, a_2) + \alpha \cdot \big[ r_2 + \gamma \cdot \max_{a_1} Q_1(s_1', a_1) - Q_2(s_2, a_2) \big] \)
        \State Set \( s_1 \gets s_1' \)
    \EndWhile
\EndFor
\State \Return Trained Q-functions \( Q_1 \), \( Q_2 \)
\end{algorithmic}
\end{algorithm}

\begin{figure*}[h!]
    \centering
    \includegraphics[width=1.0\textwidth, height=10cm]{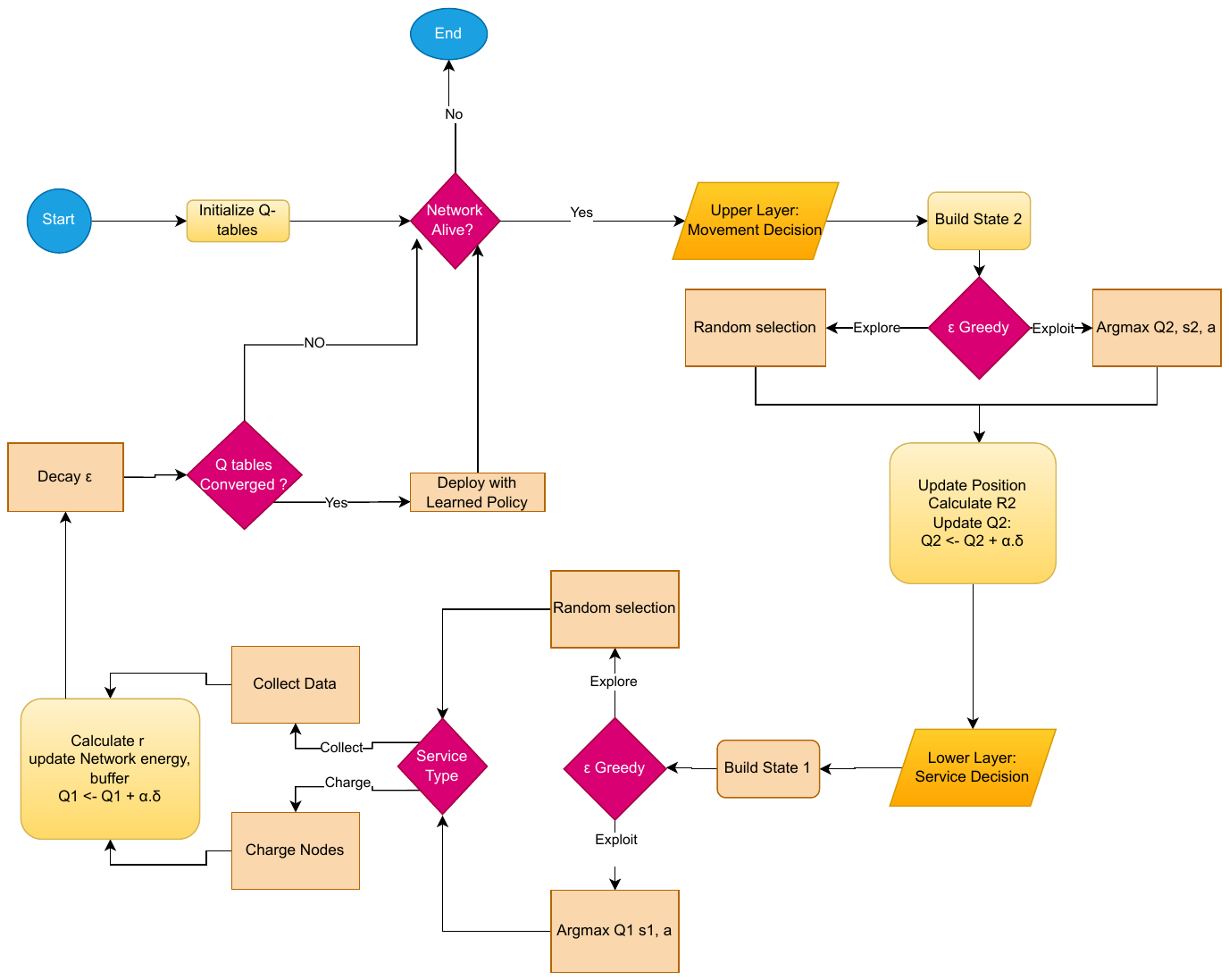} 
    \caption{Hierarchical decision-making flow of DCR-HRL. Once the Q-tables are initialized, the algorithm loops until the end of network life. In the higher level, the agent constructs State 2 and picks a moving direction using either $\epsilon$-greedy exploration (random choice) or exploitation ($\arg\max Q_2$). It moves, calculates reward $R_2$, and updates $Q_2$. In the lower level, the agent constructs State 1 and chooses between charging and collecting data by using $\epsilon$-greedy exploration or exploitation ($\arg\max Q_1$). It performs the chosen action, calculates reward $r$, updates energy and buffer levels of the network, and updates $Q_1$. After every iteration, $\epsilon$ decays. The loop checks whether the Q-tables have converged; if no, the loop starts again from checking the network status; otherwise, it deploys the agent with the learned policy.}
    \label{fig:workflow}
\end{figure*}

To clarify this procedure, Figure \ref{fig:workflow} illustrates the hierarchical workflow of the agent, showing how movement and service decisions are sequentially executed under the two-layer structure.

\section{Performance Evaluation}

In this section, we introduce our experimental evaluation of the proposed DCR-HRL approach. We conduct comprehensive simulations to assess the performance of our hierarchical reinforcement learning framework across various network conditions and compare it against state-of-the-art mobile agent strategies for wireless sensor networks.

\subsection{Simulation Setup}

The experimental evaluation is conducted using a Python 3.10 simulation environment running on Google Colab with T4 GPU acceleration. The network topology consists of a 100×100 meter deployment area divided into a 5×5 grid structure, where each 20×20 meter cell serves as a potential service location for the mobile agent. Sensor nodes are uniformly distributed across the network area using random placement within grid boundaries to ensure balanced coverage. The simulation implements realistic energy consumption models, wireless communication protocols, and mobile agent operational constraints as specified in the system model. Each experimental scenario is executed multiple times with different random seeds to ensure statistical reliability of the results.

\subsubsection{Compared Methods}

We compare our work with the following approaches:
\begin{itemize}
    \item \textbf{Sequential Baseline \cite{s24185931}:} A deterministic sequential scheduling policy where the mobile agent follows predefined movement patterns and alternates between charging and data collection services without learning or adaptation.
    \item \textbf{Multi-Agent \cite{GOYAL2024103943}:} A coordinated approach using two mobile agents with spatial division, where one agent specializes in charging operations while the other focuses on data collection within designated network regions.
    \item \textbf{DQN-MAS \cite{JIANG2024123752}:} A state-of-the-art Deep Q-Network Mobile Agent with Priority Service that uses single-layer reinforcement learning for combined movement and service decisions.
    \item \textbf{Collect-Only:} A single-task baseline in which the mobile agent only performs data collection.
    \item \textbf{Charge-Only:} A single-task baseline in which the mobile agent only performs wireless charging operations.
    \item \textbf{MDDRL-JSSTC \cite{app13148513}:} A deep reinforcement learning charging baseline using a multi-discrete-action Deep Q-Network with a bidirectional GRU to jointly select the charging destination and an adaptive charging threshold.
    \item \textbf{DCR-HRL:} Our proposed hierarchical reinforcement learning approach.
\end{itemize}

DCR-HRL and DQN-MAS differ fundamentally in their internal state representation and action-space structure: DCR-HRL decomposes decision-making into two coordinated tabular Q-tables operating over discretized, layer-specific states, whereas DQN-MAS employs a single deep network that jointly represents movement and service decisions in one unified state-action space. This divergence is not incidental but reflects the central research question of this paper—whether hierarchical decomposition of the service-selection and movement-planning sub-problems offers advantages over a joint, single-layer formulation—and enforcing an identical state or action space across both methods would eliminate the very design distinction under evaluation. We therefore consider the comparison valid on the basis of what is held constant rather than what is architecturally shared: both methods are trained and evaluated within the same simulation environment, network topology, energy and communication models, and mobile-agent operational constraints. Both are evaluated using identical performance metrics, including network lifetime, throughput, end-to-end delay, and energy efficiency, over the same experimental scenarios (100--500 nodes, 25--125~KB/s data rates); and both are given access to the same underlying network-state information—local and neighboring energy and buffer levels—differing only in how that information is structured and processed internally.

\subsubsection{Simulation Parameters}

The simulation parameters used in our experiments are summarized in Table~\ref{tab:simulation_params}. These parameters are carefully selected based on realistic WSN deployment scenarios and follow standard practices in mobile agent research. The network area is divided into a 5×5 grid structure to enable systematic mobile agent navigation, with each grid cell representing a service subdomain for the mobile agent operations.

\begin{table}[htbp]
\centering
\small
\caption{Simulation Parameters}
\label{tab:simulation_params}
\begin{tabular}{|l|p{3cm}|}
\hline
\textbf{Parameter} & \textbf{Value} \\
\hline
\multicolumn{2}{|c|}{\textbf{Network Parameters}} \\
\hline
Network area & $100 \times 100$ m$^2$ \\
Number of nodes (N) & 100, 200, 300, 400, 500 \\
Grid dimensions & $5 \times 5$ \\
Initial node energy ($E_{\text{init}}$) & 20 J \\
Maximum buffer size ($B_{\text{max}}$) & 100 packets \\
Node comm. range ($R_{\text{comm}}$) & 15 m \\
Data packet size (k) & 4000 bits \\
$E_{\text{elec}}$ & 50 nJ/bit \\
$E_{\text{amp}}$ & 100 pJ/bit/m$^2$ \\
Path loss exponent (n) & 2 \\
Distance threshold ($d_{\text{thresh}}$) & 30 m \\
\hline
\multicolumn{2}{|c|}{\textbf{Mobile Agent Parameters}} \\
\hline
Initial energy ($E_{\text{MA},\text{init}}$) & 50 J \\
Movement speed & 2 m/s \\
Movement energy cost ($E_{\text{move}}$) & 0.1 J/m \\
Charging duration per grid & 8 s \\
Data collection duration & 6 s \\
Service radius & 10 m \\
Max nodes per service & 5 \\
Transfer efficiency ($\eta$) & 0.5 \\
Energy harvesting rate & 0.3 J/s \\
Agent data capacity ($B_{\text{agent}}$) & Unbounded \\
Energy threshold ($E_{\text{threshold}}$) & 10 J \\
Buffer threshold ($B_{\text{threshold}}$) & 50 packets \\
\hline
\multicolumn{2}{|c|}{\textbf{Q-Learning Parameters}} \\
\hline
Learning rate ($\alpha$) & 0.1 \\
Discount factor ($\gamma$) & 0.9 \\
Exploration rate ($\epsilon$) & $0.3 \to 0.01$ (decay) \\
State discretization & 5 bins each \\
Training episodes & 1000 \\
Episode length & No limit \\
Charging weight ($w_{\text{charge}}$) & 0.6 \\
Collection weight ($w_{\text{collect}}$) & 0.4 \\
Reward weight ($\lambda_{\text{charge}}$) & 1.0 \\
Reward weight ($\lambda_{\text{collect}}$) & 1.0 \\
Efficiency weight ($\lambda_{\text{near}}$) & 0.5 \\
Efficiency weight ($\lambda_{\text{far}}$) & 0.5 \\
Penalty weight ($\lambda_{\text{penalty}}$) & $-$0.5 \\
Movement penalty ($\lambda_{\text{move}}$) & 0.1 \\
Energy state weight ($w_1$) & 0.6 \\
Buffer state weight ($w_2$) & 0.4 \\
\hline
\end{tabular}
\end{table}

The experimental scenarios systematically vary two primary parameters: network density ranging from 100 to 500 sensor nodes to evaluate scalability, and data generation rates from 25 KB/s to 125 KB/s to assess performance under different traffic loads. These parameter variations enable comprehensive analysis of how different mobile agent strategies adapt to changing network conditions, with particular attention to congestion effects, energy distribution patterns, and service coordination efficiency across diverse operational environments.

\subsubsection{Evaluation Metrics}

Our experimental evaluation focuses on comprehensive performance assessment across multiple critical metrics that characterize wireless sensor network efficiency and longevity. We evaluate network lifetime using three key indicators: First Node Death (FND), Half Node Death (HND), and Last Node Death (LND), where we define the last node death threshold as when only 30\% of nodes remain alive, representing the point where network coverage becomes insufficient for practical applications. Network lifetime metrics measure the temporal sustainability of the sensor network under different mobile agent strategies, with FND indicating the time when energy imbalance first causes node failure, HND marking the point where half the network capacity is lost, and LND representing near-complete network degradation.

Beyond lifetime analysis, we assess operational performance through throughput, end-to-end delay, and energy efficiency metrics. Throughput quantifies the rate of successful data packet delivery from sensor nodes to the base station, measured in packets per second, reflecting the network's data gathering effectiveness under various mobile agent coordination strategies. End-to-end delay represents the time duration from packet generation at a sensor node until successful delivery to the base station, encompassing both buffering delays at intermediate nodes and mobile agent collection latencies. Energy Per Bit (EPB) measures the total network energy consumption required to successfully deliver one bit of information to the base station, providing insight into the energy efficiency of different mobile agent approaches.

\subsection{Results and Analysis}

\subsubsection{Convergence}

\begin{figure}[h!]
    \centering
    \includegraphics[width=0.50\textwidth]{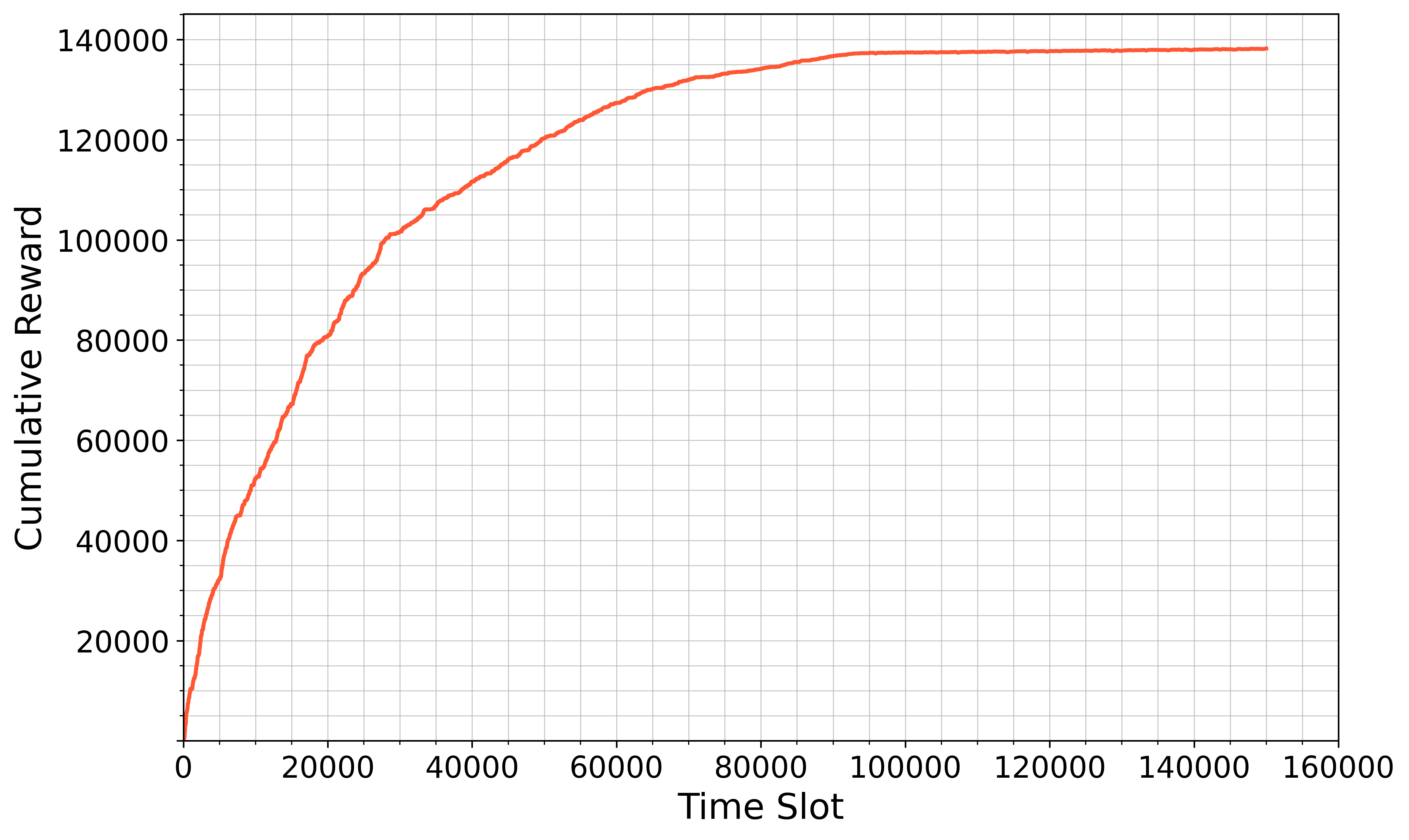} 
    \caption{Convergence of the Cumulative Reward in the Proposed Learning Approach. The reward stabilizes once the agent's policy converges, indicating that further training yields diminishing improvement.}
    \label{fig:reward_convergence}
\end{figure}

As with most learning-based approaches, our Q-learning method requires a training phase to discover an effective decision-making policy. To demonstrate the convergence speed, Figure~\ref{fig:reward_convergence} illustrates the average reward over time. This reward is the total points the mobile agent earns during a simulation by successfully collecting data, charging sensor nodes, and covering the network area. By examining the cumulative reward curve across several simulation runs, we can determine when the agent's learning stabilizes. If the reward stops significantly increasing, it indicates that the agent has already found the optimal decision strategy, and further training won’t improve its performance very much.

\subsubsection{Discretization Sensitivity Analysis}
To justify the use of five bins, a sensitivity analysis was performed by varying the discretization granularity (3, 4, 5, 6, 7, and 10 bins) and measuring its effect on convergence speed, network lifetime (FND), and the final per-episode Q-value change ($|\Delta Q|$). As shown in Table~\ref{tab:bin_sensitivity}, coarse discretizations with 3 and 4 bins fail to stabilize within the training budget, converging only at episodes 996 and 993, respectively, because too few bins collapse distinct network states into the same table entry and prevent the agent from learning a consistent policy. Five bins converge fastest, at episode 10, whereas 6, 7, and 10 bins require progressively more episodes (24, 30, and 52) as their larger state spaces take longer to sample adequately, despite inflating the Q-table size by up to $33\times$. Network lifetime is insensitive to granularity: FND stays within a narrow band across the converged configurations, confirming that finer discretization yields no lifetime benefit to offset its added cost. The steadily decreasing $|\Delta Q|$ at higher bin counts reflects the smaller average per-cell update within a larger, more sparsely visited table, rather than any improvement in policy quality. Five bins therefore offers the best trade-off between convergence speed, representational accuracy, and computational tractability at our network scale.

\begin{table}[h!]
\centering
\caption{Sensitivity Analysis: Effect of State Space Discretization on 
Learning Performance}
\label{tab:bin_sensitivity}
\begin{tabular}{|c|c|c|c|c|}
\hline
\textbf{Bins} & \textbf{Q-table Size} & \textbf{Conv.\ Ep.\textsuperscript{b}} & 
\textbf{Avg FND (s)} & \textbf{Per-ep. $|\Delta Q|$} \\
\hline
3  & 3,150      & 996 & 408.2 & 0.00420 \\
4  & 7,200     & 993 & 442.4 & 0.00127 \\
\textbf{5}  & \textbf{13,750}    & \textbf{10}  & \textbf{397.0} & \textbf{0.00060} \\
6  & 23,400    & 24  & 373.2 & 0.00029 \\
7  & 36,750    & 30  & 365.1 & 0.00016 \\
10 & 105,000  & 52  & 372.9 & 0.00005 \\
\hline
\multicolumn{5}{l}{\footnotesize \textsuperscript{b}Reward-stabilized at 90\% of final value.}\\
\end{tabular}
\end{table}

\subsubsection{Comparison with Baseline Computational Cost}
To address the complexity and computation-time analysis, Table~\ref{tab:complexity_compare} compares DCR-HRL against DQN-MAS, Multi-Agent, MDDRL-JSSTC and Sequential under matched training and environment conditions. DCR-HRL's Q-tables scale as $O(G \cdot b^3)$ and $O(G \cdot b^2)$ with grid count $G$ and bin count $b$, but inference remains $O(1)$ regardless of $b$; DQN-MAS instead uses a fixed-size network with $O(d \cdot h)$ forward-pass cost, trading storage growth for a fixed parametric cost at the expense of stability. DCR-HRL trains faster (78.0~s vs. 94.8~s) and converges in far fewer episodes (10 vs. 176) than DQN-MAS, reflecting the provable convergence guarantees of tabular Q-learning versus a bootstrapped single network. Its inference is slower (22.4~$\mu$s vs. 4.5~$\mu$s) and its Q-tables use roughly 27$\times$ more memory (110~KB vs. 4.05~KB)---an explicit storage-for-stability trade-off---though both remain negligible next to the 13-second decision cycle and the non-adaptive rule-based baselines (Multi-Agent 0.10~$\mu$s, Sequential 0.08~$\mu$s). MDDRL-JSSTC's Bi-GRU architecture converges in 58 episodes, with a training time of 138.7~s, inference latency of 313.8~$\mu$s, and memory footprint of 146.8~KB.

\begin{table}[h!]
\centering
\caption{Training, Inference, and Memory Comparison Across Methods}
\label{tab:complexity_compare}
\resizebox{\columnwidth}{!}{%
\begin{tabular}{|l|c|c|c|c|}
\hline
\textbf{Method} & \textbf{Train Time} & \textbf{Conv.\ Ep.\textsuperscript{b}} & \textbf{Inference} & \textbf{Memory} \\
\hline
DCR-HRL & 78.0 s & 10 & 22.4 $\mu$s & 110.0 KB \\
DQN-MAS & 94.8 s & 176 & 4.5 $\mu$s & 4.05 KB \\
MDDRL-JSSTC & 138.7 s & 58 & 313.8 $\mu$s & 146.8 KB \\
Multi-Agent & N/A & N/A & 0.10 $\mu$s & $<$1 KB \\
Sequential & N/A & N/A & 0.08 $\mu$s & $<$1 KB \\
\hline
\end{tabular}%
}
\begin{tabular}{l}
\multicolumn{1}{l}{\footnotesize \textsuperscript{b}Reward-stabilized at 90\% of final value.}
\end{tabular}
\end{table}

\subsubsection{Scalability Beyond the 5$\times$5 Setting} 
For a grid of $G$ cells and $b$ bins per state variable, the upper-layer table holds $G \cdot b^3 \cdot |A_2|$ entries and the lower-layer table $G \cdot b^2 \cdot |A_1|$, so total storage scales as $O(G \cdot b^3)$, while inference remains $O(1)$ irrespective of $G$ and $b$, since action selection requires only a fixed number of table lookups. To verify this empirically, we re-trained DCR-HRL on grid sizes from 7$\times$7 to 20$\times$20 (up to 16$\times$ the original grid count) and node counts from 500 to 2500, averaging results over three random seeds per configuration. Table~\ref{tab:scalability} reports Q-table size, memory footprint, convergence episodes, training time, and average FND for each configuration. Q-table size grows from 26,950 to 220,000 entries as grid dimension increases, matching the predicted scaling. Convergence episodes increase only moderately across this range, from 23.0 at 7$\times$7 to 60.0 at 20$\times$20, and training time grows sublinearly within the sweep, rising by roughly 10\% (111.9~s to 123.2~s) despite an 8$\times$ increase in state-space size, since the per-epoch cost of a table lookup and update is independent of table size and the episode budget is held fixed. Tabular Q-learning therefore remains tractable well beyond the originally evaluated scale. Within each grid size, increasing node count consistently reduces convergence episodes, since denser node placement fills buffers past the collection threshold more quickly and produces a stronger, more consistent reward signal for the agent to stabilize around. Average FND declines with grid size, from 233.4~s at 7$\times$7 with 500 nodes to 101.5~s at 20$\times$20 with 2500 nodes; this reflects the single mobile agent needing substantially longer to revisit any given cell as grid size grows, a physical coverage limitation already discussed in Section~\ref{sec:scalability_considerations}.

Based on the operating regimes evaluated in this study, we consider the $5 \times 5$ grid with 100--500 sensor nodes as the recommended operating range for the single-agent architecture. Beyond this range, the scalability results in Table 5 indicate that spatial coverage becomes the main limitation. In particular, for 1,000 nodes, increasing the grid size from $7 \times 7$ to $10 \times 10$ reduces the average FND from 230.9 s to 115.3 s. Across the full scalability evaluation, the average FND decreases from 233.4 s at $7 \times 7$ with 500 nodes to 101.5 s at $20 \times 20$ with 2,500 nodes. This degradation reflects the longer revisit time required by a single mobile agent as the deployment area expands.

\begin{table*}[h!]
\centering
\small
\caption{Scalability Analysis: Q-table Size, Convergence, and Performance Across Grid Sizes and Node Counts (averaged over 3 seeds)}
\label{tab:scalability}
\begin{tabular}{|c|c|c|c|c|c|c|}
\hline
\textbf{Grid} & \textbf{Nodes} & \textbf{Q-table} & \textbf{Mem.} & \textbf{Conv.} & \textbf{Train} & \textbf{Avg} \\
 & & \textbf{Entries} & \textbf{(KB)} & \textbf{Ep.} & \textbf{Time (s)} & \textbf{FND (s)} \\
\hline
7$\times$7   & 500  & 26,950  & 215.6   & 23.0 & 111.9 & 233.4 \\
7$\times$7   & 1000 & 26,950  & 215.6   & 15.7 & 112.2 & 230.9 \\
10$\times$10 & 1000 & 55,000  & 440.0   & 26.0 & 113.3 & 115.3 \\
10$\times$10 & 1500 & 55,000  & 440.0   & 17.0 & 113.3 & 112.4 \\
15$\times$15 & 1500 & 123,750 & 990.0   & 47.7 & 118.3 & 107.0 \\
15$\times$15 & 2000 & 123,750 & 990.0   & 43.7 & 118.8 & 106.5 \\
20$\times$20 & 2000 & 220,000 & 1760.0  & 60.0 & 123.2 & 102.4 \\
20$\times$20 & 2500 & 220,000 & 1760.0  & 48.3 & 121.2 & 101.5 \\
\hline
\end{tabular}
\end{table*}

\subsubsection{Energy Harvesting Rate Sensitivity Analysis}
To empirically validate the stochastic harvesting process introduced in the mobile agent energy update, we performed a sensitivity analysis sweeping the mean harvesting rate from 0.0 to 1.2 J/s, holding the 5$\times$5 grid, 300-node configuration, and learned policy fixed so that harvesting rate is the only varying factor, with results averaged over three random seeds and summarized in Table~\ref{tab:harvest_sensitivity}. Within the tested range, mean harvesting rate has no detectable effect on FND specifically, because early-lifetime node deaths are dominated by forwarding-load imbalance near the base station that manifests within the first several decision epochs, before harvesting-rate differences can meaningfully affect which grids the agent has serviced. Harvesting rate instead primarily affects the agent's own operational sustainability, reflected in the critical-low energy fraction and skipped-service count, and, through that channel, the later-stage network degradation captured by HND and LND: below the paper's default rate of 0.3 J/s, the agent's battery frequently falls under the critical floor needed to complete a service action (critical-low fractions of 0.483, 0.464, and 0.294 at 0.0, 0.1, and 0.2 J/s, respectively), forcing it to skip charging or collection outright (80--108 skipped actions on average), whereas at 0.3 J/s and above, the harvested energy sustains continuous operation and both diagnostics drop to zero. Correspondingly, HND and LND improve as the harvesting rate rises from 0.0 to 0.3 J/s, from 6991.7~s to 7228.0~s and from 9848.2~s to 10473.9~s respectively; this confirms that 0.3 J/s is the rate at which sustained agent operation is achieved, with only marginal further gains at higher rates.

\begin{table}[h!]
\centering
\caption{Sensitivity Analysis: Effect of Mean Energy Harvesting Rate on Network Performance}
\label{tab:harvest_sensitivity}
\setlength{\tabcolsep}{3pt}
\renewcommand{\arraystretch}{1.1}
\begin{tabular}{|c|c|c|c|c|c|}
\hline
\textbf{Rate} & \textbf{FND} & \textbf{HND} & \textbf{LND} & \textbf{Crit.} & \textbf{Skip.} \\
\textbf{(J/s)} & \textbf{(s)} & \textbf{(s)} & \textbf{(s)} & \textbf{Low} & \textbf{Act.} \\
\hline
0.0 & 356.2 & 6991.7 & 9848.2 & 0.483 & 80.1 \\
0.1 & 364.3 & 7061.7 & 9904.2 & 0.464 & 107.4 \\
0.2 & 343.8 & 7123.0 & 10169.4 & 0.294 & 108.0 \\
\textbf{0.3} & \textbf{365.5} & \textbf{7228.0} & \textbf{10473.9} & \textbf{0.000} & \textbf{0.0} \\
0.5 & 347.0 & 7292.4 & 10546.0 & 0.000 & 0.0 \\
0.8 & 365.0 & 7314.1 & 10702.0 & 0.000 & 0.0 \\
1.2 & 376.7 & 7541.5 & 10973.9 & 0.000 & 0.0 \\
\hline
\end{tabular}
\end{table}

\subsubsection{Empirical Validation of the Single-Antenna Constraint}
We evaluate the SINR at the mobile agent's data receiver under simultaneous versus sequential operation. Since data collection is broadcast-based, a responding node radiates a single fixed setting of $P_{\mathrm{tx}} = 0.80$~mW ($-0.97$~dBm) sized to reach the service boundary, with the received power decaying as $(d_0/d)^{n}$ over the distance $d$ to the agent. In simultaneous mode the co-located charging transmitter leaks into the data receiver as self-interference, attenuated only by the front-end isolation $\kappa = 30$~dB; under sequential operation no charging signal is present and the interference term vanishes. The resulting SINR is:
\begin{equation}
\mathrm{SINR}(d) = \frac{P_{\mathrm{tx}} \left(d_0/d\right)^{n}}{P_{\mathrm{charge}} \cdot 10^{-\kappa/10} + N_0}
\label{eq:sinr_model}
\end{equation}
where $P_{\mathrm{tx}}$ is the node's radiated data power, $d_0 = 1$~m is the reference distance, $n$ is the path loss exponent, $P_{\mathrm{charge}} = 1.5$~J/s ($=1.5$~W) is the charging power radiated by the agent, $\kappa$ is the front-end isolation between the agent's charging transmitter and its data receiver, and $N_0 = 10^{-13}$~W ($-100$~dBm) is the receiver noise floor.
Table~\ref{tab:sinr_validation} reports the SINR at four distances spanning the service radius. Sequential operation yields 79--99~dB against $-22.7$ to $-2.7$~dB for simultaneous operation, a constant 101.8~dB penalty set by the leakage-to-noise ratio alone. Simultaneous reception at the service boundary would require 58.7~dB of isolation. A packet-level Monte Carlo evaluation over 30 random seeds, in which the agent visits all 25 subdomains and serves up to $M_{\mathrm{max}}$ in-range nodes drawn uniformly per visit, yields a packet delivery ratio of 0.45\% under simultaneous operation at a 0~dB threshold (standard deviation 0.73\%, with no successful reception in 20 of the 30 seeds), confined to nodes within 0.73~m of the agent, against error-free reception under sequential operation. At a 6~dB decoding threshold, the simultaneous delivery ratio falls to zero across all seeds.
\begin{table}[h!]
\centering
\caption{SINR at the Mobile Agent's Data Receiver: Sequential vs. Simultaneous Operation ($\kappa = 30$~dB)}
\label{tab:sinr_validation}
\begin{tabular}{|c|c|c|c|}
\hline
\textbf{$d$ (m)} & \textbf{Mode} & \textbf{SINR (dB)} & \textbf{Reception} \\
\hline
\multirow{2}{*}{1} & Sequential   & 99.0 & Success \\
\cline{2-4}
                    & Simultaneous & $-2.7$ & Fails \\
\hline
\multirow{2}{*}{2} & Sequential   & 93.0 & Success \\
\cline{2-4}
                    & Simultaneous & $-8.8$ & Fails \\
\hline
\multirow{2}{*}{5} & Sequential   & 85.1 & Success \\
\cline{2-4}
                    & Simultaneous & $-16.7$ & Fails \\
\hline
\multirow{2}{*}{10 ($R_{\mathrm{service}}$)} & Sequential & 79.0 & Success \\
\cline{2-4}
                                     & Simultaneous & $-22.7$ & Fails \\
\hline
\end{tabular}
\end{table}

\subsubsection{Network Lifetime}

Network lifetime is evaluated using three key metrics: FND, HND, and LND. FND measures the time until the first sensor depletes its energy and fails, indicating when energy imbalance begins to affect network operation. HND represents the point when 50\% of nodes have failed, marking significant network degradation. LND indicates near-complete network failure when only 30\% of nodes remain operational. All results are averaged over $N=10$ independent random seeds and reported as mean $\pm$ standard deviation, with 95\% confidence intervals computed as $\bar{x} \pm t_{0.975,9}\cdot s/\sqrt{10}$. FND shows the largest seed-to-seed variation (coefficient of variation 5--15\%) since it depends on the single earliest node failure, whereas HND and LND, being aggregate milestones over many nodes, are comparatively stable across seeds (coefficient of variation 1--2\%).

\begin{figure*}[h!]
    \centering
    {\includegraphics[width=1.0\textwidth]{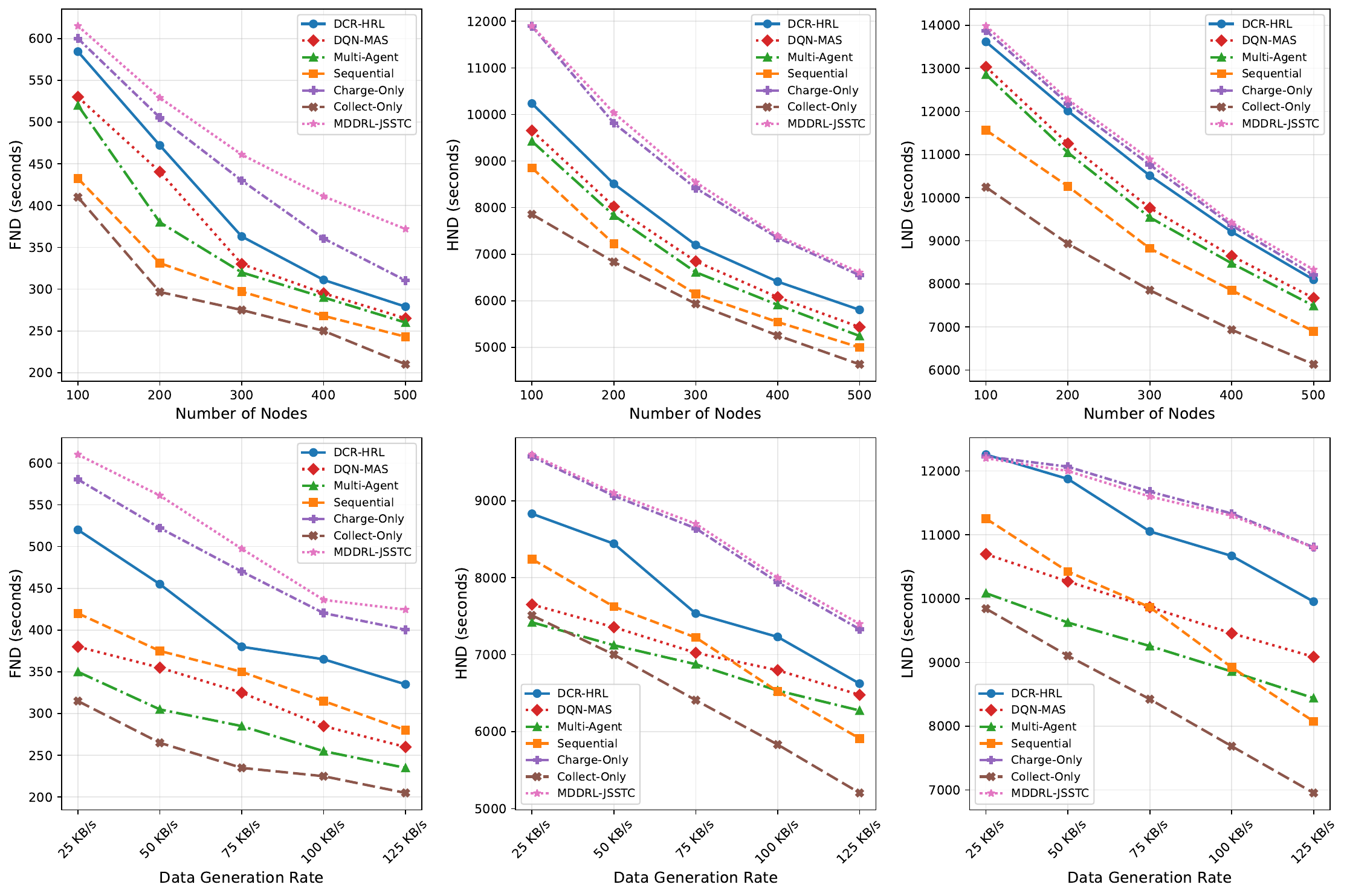}}
    \caption{Network lifetime comparison (FND, HND, and LND) across varying node densities and data generation rates for all evaluated methods.}
    \label{fig:lifetime}
\end{figure*}

\textbf{Impact of Node Density:} As shown in the top row of Figure~\ref{fig:lifetime}, increasing node density from 100 to 500 nodes reduces network lifetime across all metrics. This occurs because denser networks generate more data, and nodes near the base station must forward traffic from more neighbors, accelerating their energy depletion. DCR-HRL consistently outperforms Sequential, Multi-Agent, DQN-MAS, and Collect-Only by 5--40\% in FND across all density levels ($584.3 \pm 48$ s at 100 nodes to $279.0 \pm 41$ s at 500 nodes). The hierarchical decision-making intelligently balances charging and data collection, preventing premature energy depletion in critical bottleneck nodes near the base station. Charge-Only ($600.0 \pm 38$ to $310.3 \pm 44$) and MDDRL-JSSTC ($615.0 \pm 45$ to $372.0 \pm 42$) remain close to each other at lower densities, with MDDRL-JSSTC pulling ahead as density increases, as its deep-RL-based adaptive charging scheduler outperforms Charge-Only. The single-task baselines reveal the importance of joint optimization: Charge-Only and MDDRL-JSSTC achieve the highest FND by dedicating the agent's time to energy replenishment, but neither delivers data directly, while Collect-Only records the lowest FND among all methods as unreplenished nodes near the base station deplete rapidly under heavy forwarding loads.

\textbf{Impact of Data Generation Rate:} The bottom row of Figure~\ref{fig:lifetime} shows performance under varying traffic loads (25-125 KB/s). Higher data rates accelerate energy consumption due to increased transmission and forwarding activity. DCR-HRL consistently outperforms Sequential, Multi-Agent, DQN-MAS, and Collect-Only across all traffic loads ($520.0 \pm 32$ s at 25 KB/s to $335.0 \pm 12$ s at 125 KB/s). The lower-layer agent adaptively prioritizes charging energy-critical nodes before they fail, while the upper-layer agent optimizes movement to reach high-demand areas efficiently. Charge-Only ($580.3 \pm 29$ to $400.4 \pm 21$) and MDDRL-JSSTC ($610.1 \pm 35$ to $424.4 \pm 57$) again lead FND at every rate, with MDDRL-JSSTC's adaptive charging scheduler maintaining a consistent edge over Charge-Only. Multi-Agent and DQN-MAS show intermediate performance—Multi-Agent benefits from workload division but lacks adaptive coordination, while DQN-MAS struggles with the coupled optimization of both movement and service decisions in a single-layer framework. Collect-Only degrades most severely under higher traffic loads since unreplenished relay nodes exhaust their energy faster as data generation increases, recording the lowest FND at every rate, whereas Charge-Only and MDDRL-JSSTC maintain the highest lifetime at the cost of providing no direct data delivery to the base station.

\subsubsection{End-to-End Delay}
Average end-to-end delay measures the time from packet generation at a sensor node until delivery to the base station, expressed in milliseconds. Lower delay indicates more responsive data delivery. All results are averaged over $N=10$ independent random seeds and reported as mean $\pm$ standard deviation, with 95\% confidence intervals computed as $\bar{x} \pm t_{0.975,9}\cdot s/\sqrt{10}$.

\begin{figure*}[h!]
    \centering
    {\includegraphics[width=1.0\textwidth]{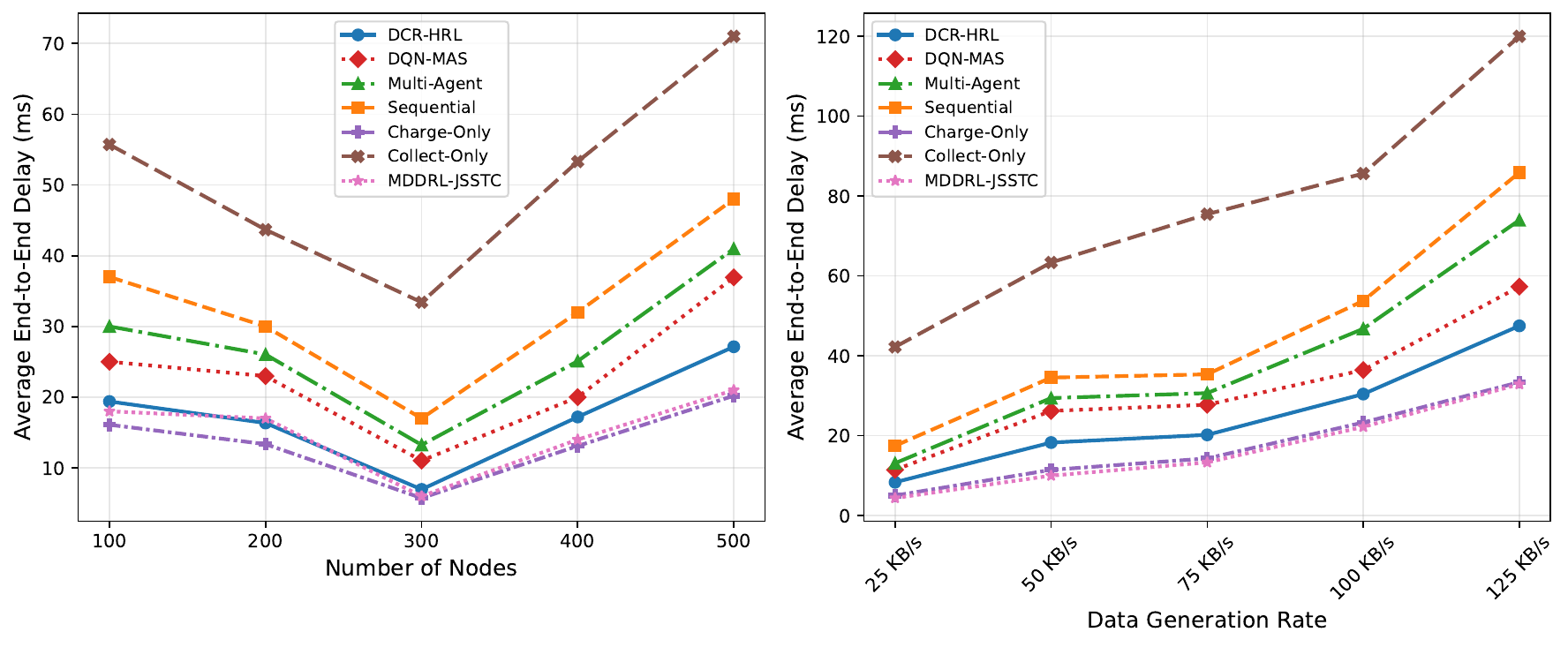} }
    \caption{Average End-to-End Delay (ms) comparison across varying node densities (left) and data generation rates (right) for all evaluated methods. Lower values indicate more responsive data delivery.}
    \label{fig:delay}
\end{figure*}

\textbf{Impact of Node Density:} The left plot of Figure~\ref{fig:delay} exhibits a characteristic U-shaped curve. At low density (100 nodes), delay is high ($19.4 \pm 0.3$ ms for DCR-HRL) because sparse networks create longer multi-hop paths with limited routing alternatives. As density increases to 300 nodes, delay decreases significantly—denser networks provide shorter paths and more efficient mobile agent coverage, with DCR-HRL achieving minimum delay of $6.9 \pm 0.5$ ms. Beyond 300 nodes, delay rises sharply as congestion dominates: channel collisions increase, buffers fill faster, and the mobile agent navigates more 
slowly through dense areas. DCR-HRL achieves a competitive delay of $27.1 \pm 0.6$ ms at 500 nodes, close to Charge-Only ($20.2 \pm 0.3$ ms) which records the lowest overall delay, because hierarchical decision-making prioritizes collecting from nearly-full buffers and charging critical nodes before failures occur. MDDRL-JSSTC remains close to Charge-Only ($6.0 \pm 0.5$--$21.0 \pm 0.5$ ms across densities). Sequential baseline shows the worst performance ($48.0 \pm 1.2$ ms) due to inflexible routing. Charge-Only achieves the lowest overall delay across all node densities because nodes are kept fully charged and can forward data immediately via multi-hop without congestion or routing failures, while Collect-Only shows the highest delay because without charging, relay nodes deplete their energy and fail, forcing data through longer and less reliable multi-hop paths.

\textbf{Impact of Data Generation Rate:} The right plot of 
Figure~\ref{fig:delay} shows delay increasing monotonically with traffic load. At 25 KB/s, DCR-HRL achieves $8.4 \pm 0.7$ ms as buffers fill slowly. Beyond 75 KB/s, delay escalates rapidly—buffers overflow, nodes deplete energy, causing path failures, and the mobile agent cannot service all areas quickly enough. DCR-HRL achieves $47.5 \pm 0.8$ ms at 125 KB/s, trailing Charge-Only ($33.4 \pm 0.3$ ms) and MDDRL-JSSTC ($32.9 \pm 0.2$ ms), both of which keep relay nodes continuously charged and avoid the path failures that inflate delay under heavy dual-task servicing. Multi-Agent degrades at high loads ($74.0 \pm 0.2$ ms) due to poor coordination, while Sequential reaches $85.9 \pm 0.3$ ms due to fixed patterns that cannot adapt to congestion. Collect-Only delay increases most steeply with traffic load since unreplenished nodes fail progressively, continuously degrading the multi-hop routing fabric and increasing path lengths to the base station, while Charge-Only and MDDRL-JSSTC remain closest to DCR-HRL as multi-hop forwarding operates continuously without node failure disruptions.

\subsubsection{Throughput}
Throughput measures the rate of successful data packet delivery from sensor nodes to the base station, expressed in packets per second (pps). This metric directly reflects the network's data gathering effectiveness under various operational conditions. Throughput is computed empirically from the simulation: for each run, we record the total number of packets successfully delivered to the base station divided by the total simulation time, summed over both mobile-agent collection and surviving multi-hop transmissions.

Throughput depends on the learned policy's emergent visitation pattern. The physical delivery mechanism admits a theoretical upper bound composed of the same two contributions, given in Equation~\eqref{eq:throughput_bound}:

\begin{equation}
Th_{\max} = \underbrace{\frac{1}{k}\cdot R_{\text{collect}} \cdot \frac{\Delta t_{\text{collect}}}{T_{\text{cycle}}}}_{\text{mobile-agent path}} + \underbrace{Th_{\text{multihop}}}_{\text{surviving multi-hop path}}
\label{eq:throughput_bound}
\end{equation}
where $k$ is the data packet size, $R_{\text{collect}}$ is the collection rate of the mobile agent, $\Delta t_{\text{collect}}$ is the duration of a collection operation, and $T_{\text{cycle}}$ is the total duration of one decision cycle. The first term is thus the agent's maximum sustainable collection rate, obtained by duty-cycling the collection rate over the fraction of each cycle spent collecting, and $Th_{\text{multihop}}$ is the throughput contributed by surviving multi-hop forwarding, given by $Th_{\text{multihop}} = N_{\text{alive}}(t) \cdot r_{\text{gen}} \cdot p_{\text{success}}$, which depends on the number of surviving nodes $N_{\text{alive}}(t)$, their per-node data generation rate $r_{\text{gen}}$, and the probability that a generated packet successfully arrives at the base station via multi-hop $p_{\text{success}} \in [0,1]$. Charging affects $Th_{\text{multihop}}$ indirectly, by keeping nodes alive and sustaining the multi-hop path. The arrival probability itself depends on buffer occupancy, residual energy, the mobile agent's current position, and other instantaneous network conditions, which is why we evaluate throughput empirically rather than through a closed-form expression alone.

\begin{figure*}[h!]
    \centering
    {\includegraphics[width=1.0\textwidth]{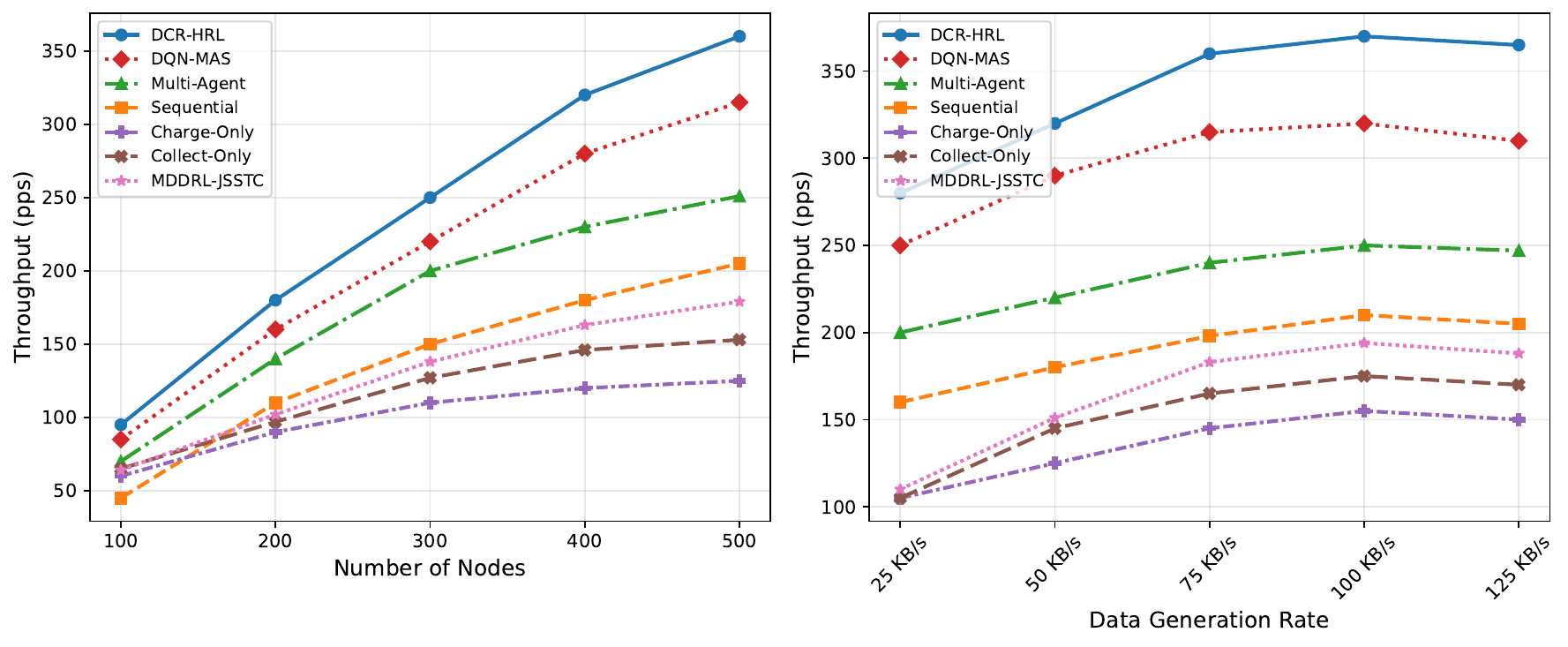}} 
    \caption{Throughput (pps) comparison across varying node densities (left) and data generation rates (right) for all evaluated methods.}
    \label{fig:throughput}
\end{figure*}

\textbf{Impact of Node Density:} In Figure~\ref{fig:throughput}, the left plot shows throughput increasing with network density, with DCR-HRL demonstrating the strongest growth trajectory. Initially, more nodes provide additional routing paths and better network connectivity, improving overall data delivery. DCR-HRL achieves the highest throughput (360 pps at 500 nodes) because the mobile agent reduces reliance on multi-hop communication—even when nodes near the base station fail, the agent continues collecting data directly from distant nodes and delivering it to the base station. This resilience to node failures is a key advantage. DQN-MAS achieves moderate performance (315 pps at 500 nodes) but suffers from the complexity of jointly optimizing movement and service decisions in a single learning framework. Multi-Agent shows slower growth (251 pps at 500 nodes), plateauing earlier due to poor coordination between the two agents, leading to redundant visits and missed coverage areas. Sequential baseline shows the lowest throughput (205 pps at 500 nodes) because its fixed patterns cannot adapt to changing network conditions or node failures. Collect-Only achieves 153 pps at 500 nodes, outperforming Charge-Only (125 pps) because the agent collects data directly at each grid visit rather than relying solely on multi-hop forwarding; however, both single-task baselines fall well below DCR-HRL, confirming that joint optimization of charging and collection is essential for high throughput. MDDRL-JSSTC trails the RL-based movement-and-service methods, reaching 179 pps at 500 nodes—above both single-task baselines despite its charging-only design providing no direct agent-side data collection, as its adaptive charging keeps more relay nodes alive to sustain multi-hop delivery than either fixed-rule baseline.

\textbf{Impact of Data Generation Rate:} In Figure~\ref{fig:throughput}, the right plot reveals saturation behavior as traffic increases. At low traffic loads (25 KB/s), DCR-HRL starts at 280 pps and increases to peak performance of 370 pps at 100 KB/s as the network becomes more active. However, beyond this point, throughput plateaus and slightly declines to 365 pps at 125 KB/s. This occurs because higher data rates cause network congestion—buffers fill faster, nodes run out of energy more quickly, and packet losses increase. DCR-HRL maintains the highest throughput across all rates because its hierarchical decision-making dynamically balances data collection from nodes with full buffers and charging of energy-depleted nodes. The adaptive prioritization prevents catastrophic failures that would completely halt data delivery. DQN-MAS shows similar saturation behavior, peaking at 320 pps before declining to 310 pps. Sequential baseline shows the poorest performance (205 pps at 125 KB/s) because it cannot respond to urgent charging or collection needs. Multi-Agent maintains relatively stable throughput (247-250 pps range at high rates) but at much lower absolute values due to coordination inefficiencies. Charge-Only records the lowest throughput across all data rates (105--150 pps) as it relies entirely on multi-hop forwarding with no direct agent collection. Collect-Only performs slightly better (105--175 pps) but approaches Sequential throughput at high data rates, demonstrating that dedicated collection without adaptive charging and scheduling cannot match the performance of DCR-HRL. MDDRL-JSSTC shows a similar pattern, peaking at 194 pps at 100 KB/s and settling around 188 pps at 125 KB/s—ahead of Charge-Only and Collect-Only.

\subsubsection{Energy Consumption}
EPB measures the total network energy consumption required to successfully deliver one bit of information to the base station, expressed in Joules per bit. Lower values indicate higher energy efficiency.

\begin{figure*}[h!]
    \centering
    {\includegraphics[width=1.0\textwidth]{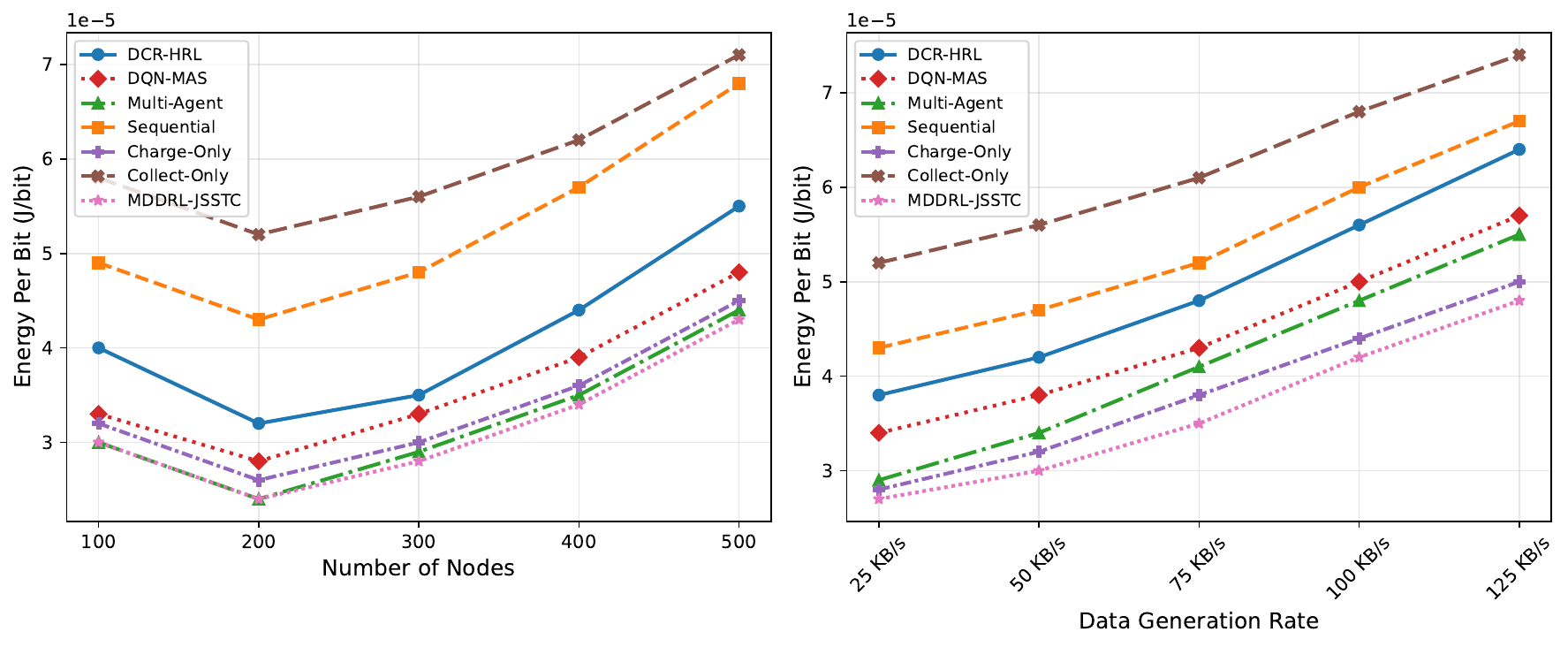}}
    \caption{Energy Per Bit (J/bit) comparison across varying node 
    densities (left) and data generation rates (right) for all evaluated methods. Lower values indicate higher energy efficiency.}
    \label{fig:energy_per_bit}
\end{figure*}

\textbf{Impact of Node Density:} The left plot of 
Figure~\ref{fig:energy_per_bit} exhibits a characteristic J-shaped curve for all approaches. At low density (100 nodes), EPB is high (4.0e-5 J/bit for DCR-HRL) because sparse networks require longer multi-hop paths—each bit traverses more hops, accumulating energy costs at each relay. As density increases to 200 nodes, EPB decreases significantly to its minimum (3.2e-5 J/bit for DCR-HRL). This improvement occurs because denser deployment creates shorter routing paths, and the mobile agent can serve multiple nearby sensors during a single visit, amortizing movement costs across more collected data. Beyond 200 nodes, EPB rises again as congestion dominates—nodes compete for channel access, causing retransmissions, increased forwarding loads deplete energy faster, and the mobile agent requires more time navigating dense areas. MDDRL-JSSTC and Multi-Agent achieve the best efficiency (2.4e-5 to 4.4e-5 J/bit, effectively tied at low density, with MDDRL-JSSTC pulling slightly ahead from 300 nodes onward) through specialized adaptive task allocation. DCR-HRL ranks below these (4.0e-5 to 5.5e-5 J/bit), deliberately trading energy efficiency for network longevity—the hierarchical approach makes aggressive service decisions to prevent node failures, consuming additional energy per bit but maintaining sustained operation. Sequential baseline shows the worst performance among dual-function methods (4.9e-5 to 6.8e-5 J/bit) due to inflexible patterns that waste energy on unnecessary movements. Charge-Only achieves competitive EPB values (3.2e-5 to 4.5e-5 J/bit), close behind Multi-Agent and MDDRL-JSSTC, because fully charged nodes avoid retransmissions caused by failed multi-hop paths. Collect-Only records the worst EPB overall (5.8e-5 to 7.1e-5 J/bit) because without charging, nodes near the base station deplete rapidly, causing path failures and excessive retransmissions that waste energy per delivered bit.

\textbf{Impact of Data Generation Rate:} The right plot of 
Figure~\ref{fig:energy_per_bit} shows EPB increasing monotonically with traffic load. At 25 KB/s, MDDRL-JSSTC achieves the best efficiency (2.7e-5 J/bit), closely followed by Charge-Only (2.8e-5 J/bit) and Multi-Agent (2.9e-5 J/bit), while DCR-HRL starts at 3.8e-5 J/bit. As traffic increases, EPB rises across all methods due to congestion effects—buffers overflow causing retransmissions, nodes exhaust energy faster requiring more frequent charging, and increased multi-hop forwarding leads to channel contention. At 125 KB/s, DCR-HRL reaches 6.4e-5 J/bit because its aggressive servicing strategy—while extending network lifetime—consumes more energy per bit to maintain connectivity under stress. MDDRL-JSSTC maintains the lowest EPB at 125 KB/s (4.8e-5 J/bit), ahead of Charge-Only (5.0e-5 J/bit) and Multi-Agent (5.5e-5 J/bit), as its adaptive charging threshold avoids over- or under-charging relative to Charge-Only's fixed rule. This trade-off reveals that optimizing for sustained network operation requires accepting higher energy consumption per bit compared to strategies focused purely on instantaneous efficiency. Collect-Only records the worst EPB (5.2e-5 to 7.4e-5 J/bit) and shows the steepest degradation with increasing traffic because higher data rates accelerate energy depletion in uncharged nodes, causing more frequent path failures and retransmissions that significantly increase the energy cost per successfully delivered bit.

\subsubsection{Joint Optimization}

Charge-Only, by dedicating all agent time to energy replenishment, achieves the highest lifetime and lowest delay by construction, since it never diverts service time to data collection; however, it delivers no direct throughput from the agent itself, relying entirely on multi-hop forwarding. DCR-HRL's contribution lies in jointly balancing both objectives: it achieves the highest throughput among all methods while remaining competitive—rather than dominant—on lifetime, delay, and energy efficiency, a trade-off no single-task baseline can offer.

\section{Conclusion and Future Work}\label{sec:conclusion}

In this study, a hierarchical reinforcement learning-based approach was proposed to improve energy efficiency and extend the lifetime of WSNs using a multi-functional mobile agent. The agent, capable of both data collection and node charging, was modeled using the MDP framework and trained with a Q-learning algorithm to make adaptive decisions based on real-time network conditions. This adaptive behavior allowed the mobile agent to optimize its movement and select the appropriate service—charging or data collection—for each area dynamically. The proposed approach aimed to reduce the burden on sensors located near the base station by distributing energy consumption more evenly throughout the network. Simulation results confirmed that the method outperforms traditional models in terms of throughput, balanced energy distribution, and network lifetime.

Although the proposed approach achieved promising results, several research directions warrant consideration. The use of more advanced reinforcement learning techniques—such as Policy Gradient methods—may improve learning efficiency and scalability in large-scale or dynamic networks. Furthermore, while the single-agent DCR-HRL framework demonstrates strong performance in small-to-medium scale networks, extending it to multi-agent deployments presents both promising opportunities and significant challenges. To address the scalability and coordination challenges identified, the hierarchical Q-learning structure of DCR-HRL could be extended to a cooperative multi-agent reinforcement learning framework, where agents share global network state information while maintaining independent Q-tables, enabling conflict-free and scalable decision-making across large and dynamic WSN environments. In addition, integration with Federated Learning (FL) systems could be investigated by adapting the reinforcement learning model to consider distributed learning requirements and node availability, potentially improving both network stability and learning accuracy. Finally, the proposed method should be evaluated under real-world conditions, including heterogeneous sensor capabilities, environmental disturbances, and hardware limitations, to assess its robustness and practical applicability in diverse deployment scenarios. Additionally, the emergence of ISAC techniques \cite{10005142} opens a promising future direction, where ISAC-enabled edge nodes could potentially assist or replace mobile agents in certain WSN scenarios by simultaneously performing data collection and network monitoring, enabling more efficient joint optimization of communication, sensing, and energy management.


\bibliographystyle{IEEEtran}
\bibliography{sn-bibliography}

\begin{IEEEbiography}[{\includegraphics[width=1in,height=1.25in,clip,keepaspectratio]{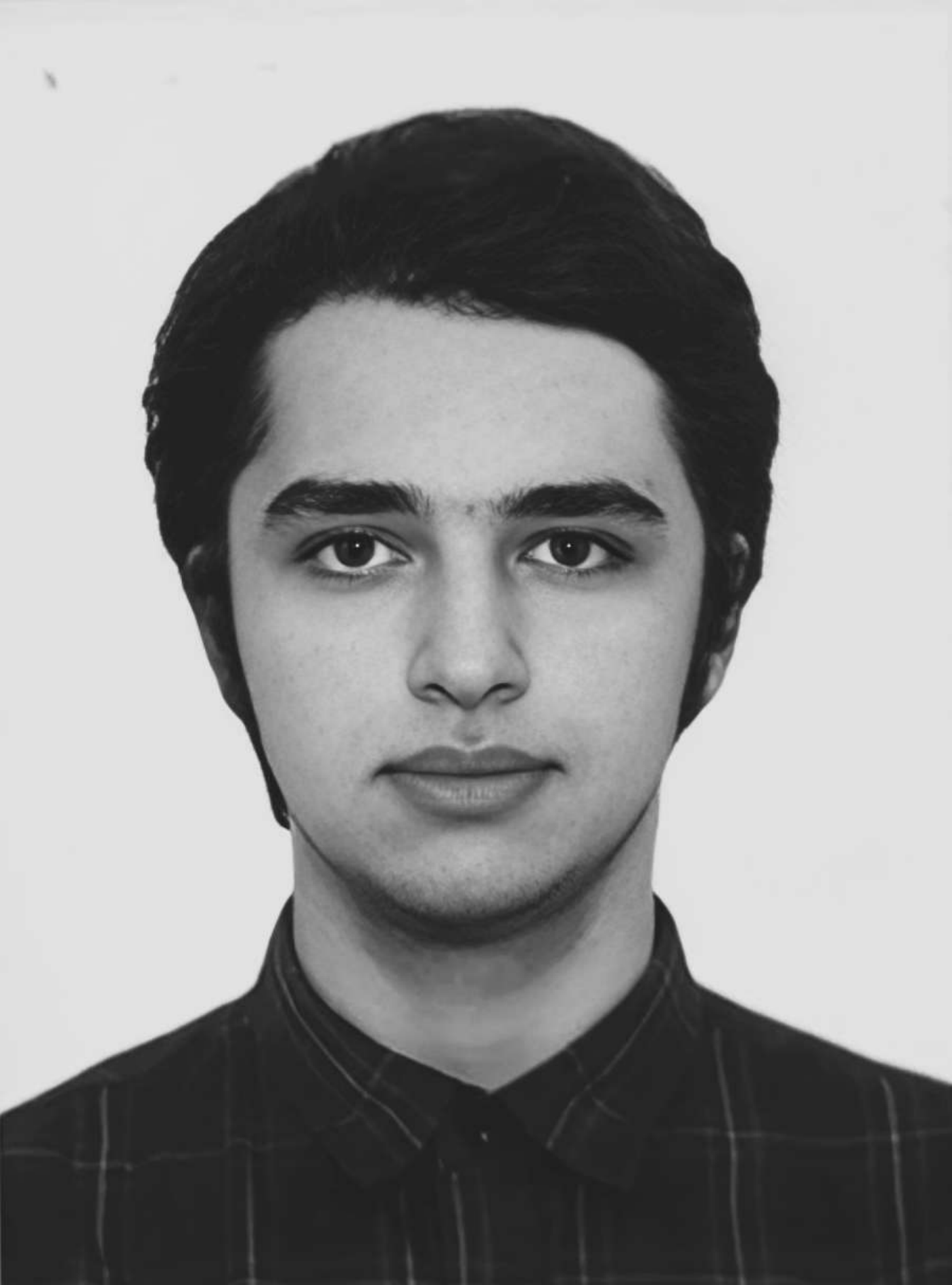}}]{ALI HEIDARIPOUR } is currently a Bachelor's student at the Computer Engineering Department, Imam Khomeini International University (IKIU), where he started in 2021. He is also a guest student at the University of Tehran (UT). His interests include Wireless Sensor Networks (WSNs), the Internet of Things (IoT), Artificial Intelligence (AI), and Quantum Computing.
\end{IEEEbiography}

\begin{IEEEbiography}[{\includegraphics[width=1in,height=1.25in,clip,keepaspectratio]{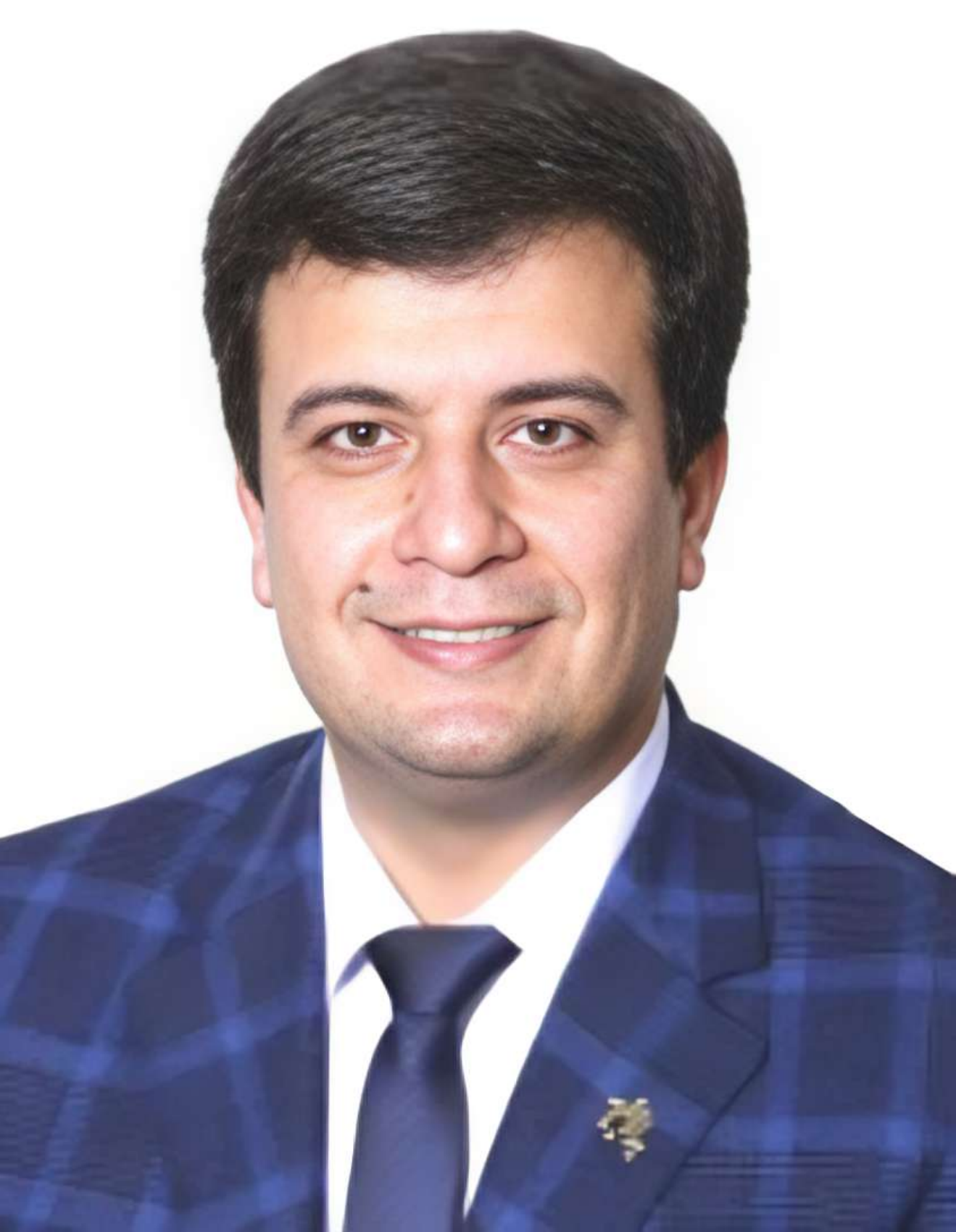}}]{NASTOOH TAHERI JAVAN } (Senior Member, IEEE) is an Assistant Professor with the Computer Engineering department at Imam Khomeini International University (IKIU), Qazvin, Iran. Dr. Taheri Javan was a post-doctoral fellow at Amirkabir University of Technology (Tehran Polytechnic), Tehran, IRAN, where he completed his M.S. and Ph.D. in computer engineering in 2007 and 2017, respectively. His research interests include wireless computer networks and network coding theory, spanning from theory, design, and implementation. He has actively collaborated with researchers in various disciplines of computer science, particularly resource management problems in the area of network architecture.
\end{IEEEbiography}

\end{document}